\documentclass[twocolumn,aps,prc,amsmath,amssymb,floatfix,nofootinbib]{revtex4-1}
\usepackage[dvips]{graphicx}
\usepackage{CJK}                     %
\usepackage{bm}                      
\usepackage{mathptmx}                
\usepackage{dcolumn}                 
\usepackage{multirow}                
\usepackage{xcolor}                  %
\usepackage[normalem]{ulem}          %
\usepackage{url}
\begin{document}

\def\bea{\begin{eqnarray}} \def\eea{\end{eqnarray}}
\def\be{\begin{equation}} \def\ee{\end{equation}}
\def\bal#1\eal{\begin{align}#1\end{align}}
\def\bse#1\ese{\begin{subequations}#1\end{subequations}}
\def\rra{\right\rangle} \def\lla{\left\langle}
\def\eps{\varepsilon}
\def\ms{M_\odot}
\def\mmax{M_\text{max}}
\def\mtov{M_\text{TOV}}
\def\esym{E_\text{sym}}
\def\pt{p_\text{th}}
\def\fm3{\;\text{fm}^{-3}}
\def\fmi{$\;{fm}^{-1}$}
\def\mfm{\;\text{MeV}\,\text{fm}^{-3}}

\long\def\new#1{\color{blue}#1\color{black}}
\title{Hot neutron stars and their equation of state}

\begin{CJK*}{UTF8}{gbsn} 

\author{Jin-Biao Wei (魏金标)$^1$}\email{jinbiao.wei@ct.infn.it}
\author{G. F. Burgio$^1$}
\author{Ad. R. Raduta$^2$}
\author{H.-J. Schulze$^1$}

\affiliation{
$^1$ INFN Sezione di Catania, Dipartimento di Fisica,
Universit\'a di Catania, Via Santa Sofia 64, 95123 Catania, Italy\\
$^2$ IFIN-HH, P.O. Box MG6, Bucharest-Magurele, Romania}

\date{\today}

\begin{abstract}
A set of microscopic,
covariant density-functional,
and non-relativistic Skyrme-type
equations of state
is employed to study the structure of purely nucleonic neutron stars
at finite temperature.
After examining the agreement with presently available
astrophysical observational constraints,
we find that the magnitude of thermal effects depends on the
nucleon effective mass as well as on the stiffness of the cold equation of state.
We evidence a fairly small but model-dependent effect of finite temperature
on stellar stability
that is correlated with the relative thermal pressure inside the star.
\end{abstract}


\maketitle

\end{CJK*}

\section{Introduction}

The recent availability of new observational data from the first
neutron star (NS) merger event GW170817 \cite{Abbott17,Abbott18}
and the Neutron Star Interior Composition Explorer (NICER) instrument
\cite{Riley19,Miller19,Riley21,Miller21}
has allowed to severly constrain the features of the {\em cold}
and {\em $\beta$-equilibrated}
equation of state (EOS) that controls the structure of NSs.
In particular, more stringent lower and upper limits on the
NS maximum mass $\mmax$ and also useful boundaries on the NS radii
have been recently established.

This has already allowed to rule out many nuclear EOSs that were
proposed in the past but do not fulfill any more the new constraints.
The most striking example is the perhaps most frequently used
microscopic EOS for theoretical NS studies, the APR EOS \cite{Akmal98},
which has been found to predict too small NS radii
in conflict with the new data.

It is therefore timely to focus on EOSs that are still compatible
with all current constraints and to examine their features and predictions for
a wider range of NS properties,
in order to be able to narrow even more the set of realistic EOSs
in the future.

This article is dedicated to such a study focussing on the finite-temperature
properties of the nuclear EOS,
which are still practically unconstrained by current astrophysical data.
However, this situation is supposed to change quickly,
in particular once features of the hot remnant of NS merger events
will be revealed by more sensitive next-generation gravitational-wave detectors
\cite{Maggiore20}.

While a large number of phenomenological and microscopic EOSs
is available for investigating cold NSs \cite{Oertel17,Burgio21},
the current subset of models extended to finite temperature
is much more limited,
which allows us to focus on only a handful of different models.
More precisely, the EOSs that fulfill our selection criteria
(discussed in detail in the following sections)
are the following:
TNTYST \cite{Togashi17} and
SRO(APR) \cite{Schneider19},
both employing the same basic two-body Argonne V18 \cite{ArgonneV18}
and the three-body Urbana UIX \cite{Carlson_NPA_1983,Pudliner_PRL_1995}
nuclear potentials and based on variational calculations
as the original APR EOS \cite{Akmal98},
the Brueckner-Hartree-Fock (BHF) EOSs V18 and N93
\cite{Li08a,Li08b,Burgio10,Lu19},
the covariant density-functional theory EOS models
Shen11 \cite{1998PThPh.100.1013S,Shen11}
(based on TM1~\cite{TM1}),
Shen20 \cite{Shen20}
(based on TM1e~\cite{TM1e}),
HS(DD2) \cite{Hempel10,DD2},
SFH(SFHx) \cite{Hempel10,Steiner13},
and FSU2H \cite{Tolos17,Tolos17b}.

We remind that differences between the original APR
and KOST \cite{Kanzawa07} interactions,
on which TNTYST \cite{Togashi17} relies,
arise from technical details in the variational procedure,
in particular simplifications in the cluster method employed by the latter.
The most striking difference regards
the presence (for APR) or absence (for KOST)
of discontinuities in the energy density as a function of density which,
in the case of APR, have been interpreted as signatures of a transition
to a pion condensate state.
Regarding the BHF EOSs V18 and N93, they include the contribution
of microscopic three-body forces consistent with the adopted two-body force;
they have a strongly repulsive character at high density,
at variance with the phenomenological Urbana three-body force (UIX)
adopted earlier \cite{Burgio10}.
In the covariant density functional
or relativistic mean field (RMF)
theory nucleons interact among each
others by the exchange of scalar-isoscalar ($\sigma$),
vector-isoscalar ($\omega$) and
vector-isovector ($\rho$) and, in some cases,
also scalar-isovector $\delta$ mesons.
In one of the considered models the strength parameters are
density dependent (DD),
while the rest employ non-linear interaction terms.
Most EOS data are publicly available in the
\textsc{CompOSE} \cite{Typel_2013} data base \cite{CompOSE}.
V18 and N93 are available in \cite{BHFSUP}.
Shen11 and Shen20 data are available also in \cite{ShenWWW}.

In addition to these, we analyze in this work also the
family of LS EOSs \cite{Lattimer91}, namely LS180, LS220, LS375.
They are based on non-relativistic mean-field Skyrme-like effective
interactions and are publicly available in \cite{LSWWW}.
Although they do not fulfill our selection citeria,
their particular way of construction
allows one to investigate very clearly certain temperature effects.

As currently the finite-temperature sector of these EOSs
is practically unaffected by observational constraints,
they exhibit rather varying thermal behaviors.
In particular, it has been noted
\cite{Kaplan14,Lu19,Schneider20,Raduta20}
that the predictions for the effects of temperature on stellar stability
are widely varying:
RMF and Skyrme-type models usually predict
increasing stability (maximum gravitational mass) with temperature
\cite{Baumgarte95,Prakash97,Kaplan14},
whereas BHF results \cite{Nicotra06,Li10,Burgio10,Lu19,Figura20,Figura21}
indicate in general a slight reduction of the maximum mass.
We will try to analyze in some detail this phenomenon
and try to establish a correlation with features of the EOS.

Our paper is organized as follows.
The general formalism of the finite-temperature NS EOS
is briefly reviewed in Sec.~\ref{s:gen}.
The properties of the cold EOSs and their observational constraints
and then the extension to finite temperature
and an analysis of stellar stability
are presented in Sec.~\ref{s:res}.
Conclusions are drawn in Sec.~\ref{s:end}.

\section{Formalism}
\label{s:gen}

We do not enter into the details of how finite temperature is introduced in
the various models,
which can be found in the respective original publications,
but analyze in a pragmatic way the effect of finite temperature
for different physical quantities.

In general,
the free energy density of hot stellar matter consists of three contributions,
\be
 f = f_N + f_L + f_\gamma \:,
\label{e:f}
\ee
where $f_N$ is the nucleonic part;
$f_L$ denotes the contribution of leptons ($e,\mu,\nu_e,\nu_\mu$,
and their antiparticles);
$f_\gamma$ stands for the contribution of the photon gas.
All thermodynamical quantities are computed
in a consistent way from the total free energy density $f$
as function of partial number densities $\rho_i$
and temperature $T$,
namely one obtains chemical potentials~$\mu_i$,
pressure~$p$, internal energy density~$\eps$,
and entropy density~$s$ as
\bea
 \mu_i &=& \frac{\partial f}{\partial \rho_i}
 \Big|_{T,\{\rho_j\}_{j \neq i}} \:,
\\
 p &=& \rho^2 {\partial{(f/\rho)}\over \partial{\rho}}\Big|_{T}
 = \sum_i \mu_i \rho_i - f \:,
\label{e:eosp}
\\
 \eps &=& f + Ts \:,\quad
 s = -{{\partial f}\over{\partial T}}\Big |_{\{\rho_i\}} \:,
\label{e:eose}
\eea
where $\rho_i$ stands for the density of the species $i=n,p$ and
$\rho=\rho_n+\rho_p$ is the baryon number density.
All these quantities are directly affected by finite temperature.

However, finite temperature also modifies indirectly
the composition of stellar matter
governed by the weak interaction.
In this article we study exclusively $\beta$-stable
and neutrino-free nuclear matter,
relevant for physical systems evolving slow enough to justify these assumptions.
In $\beta$-stable nuclear matter
the chemical potential of any particle $i=n,p,l$ is uniquely determined
by the conserved quantities baryon number $B_i$, electric charge $Q_i$,
and weak charges (lepton numbers) $L^{(e)}_i$, $L^{(\mu)}_i$:
\be
 \mu_i = B_i\mu_B + Q_i\mu_Q
 + L^{(e)}_i\mu_{\nu_e}  + L^{(\mu)}_i\mu_{\nu_\mu} \:.
\label{mufre:eps}
\ee
For stellar matter containing nucleons and leptons as relevant degrees
of freedom,
the chemical equilibrium conditions read explicitly
\be
 \mu_n - \mu_p = \mu_e = \mu_\mu \:.
\label{beta:eps}
\ee
At given baryon density $\rho$,
these equations have to be solved together with the
charge-neutrality condition
\be
 \sum_i Q_i \rho_i = 0 \:.
\label{neutral:eps}
\ee
Using the hadronic and leptonic chemical potentials,
one can calculate the composition of $\beta$-stable stellar matter,
and then the total pressure $p$ and the internal energy density $\eps$,
through the thermodynamical relations
expressed by Eqs.~(\ref{e:eosp},\ref{e:eose}).
Once the EOS $p(\eps)$ is specified,
the stable configurations of a nonrotating spherically-symmetric NS
can be obtained from the well-known hydrostatic equilibrium equations
of Tolman, Oppenheimer, and Volkoff \cite{Shapiro08}
for pressure $p(r)$,
enclosed gravitational mass $m(r)$,
and baryonic mass $m_B(r)$
\bea
 {dp\over dr} &=& -\frac{Gm\eps}{r^2}
 \frac{\big( 1+ p/\eps \big) \big( 1 + 4\pi r^3p/m \big)}
 {1-2Gm/r} \:,
\\
 \frac{dm}{dr} &=& 4\pi r^2\eps \:,
\\
 \frac{dm_B}{dr} &=& 4\pi r^2 \frac{\rho m_N}{\sqrt{1-2Gm/r}}  \:,
\label{e:tov}
\eea
where $m_N=1.67\times 10^{-24}$g is the nucleon mass
and $G= 6.67408 \times 10^{-8}\text{cm}^3\text{g}^{-1}\text{s}^{-2}$
the gravitational constant.
For a chosen central value of the energy density,
the numerical integration of these equations provides the
mass ($M,M_B$) -- central density ($\rho_c$) or radius ($R$) relations.

The solution of these equations depends obviously on the
temperature profile $T(r)$.
In any realistic simulation of an astrophysical scenario at finite temperature
(supernova explosion, proto NS formation, binary NS merger),
the TOV equations are therefore embedded in a detailed
and self-consistent dynamical simulation of the temperature evolution.
We do not perform such detailed studies here,
but our current aim is just to identify the global effect of
finite temperature on the stellar stability,
as motivated in the Introduction.
We therefore focus in the following on the frequently used isentropic
temperature profile with $S/A=s/\rho=2$,
which involves typical temperatures of about 50 MeV.

\section{Results}
\label{s:res}

In the following we present the results of our numerical calculations
regarding the properties of cold and hot NS matter and the structure of NSs.

\subsection{Cold neutron stars}

\begin{figure}[t]
\vspace{-12mm}
\centerline{\includegraphics[scale=0.34]{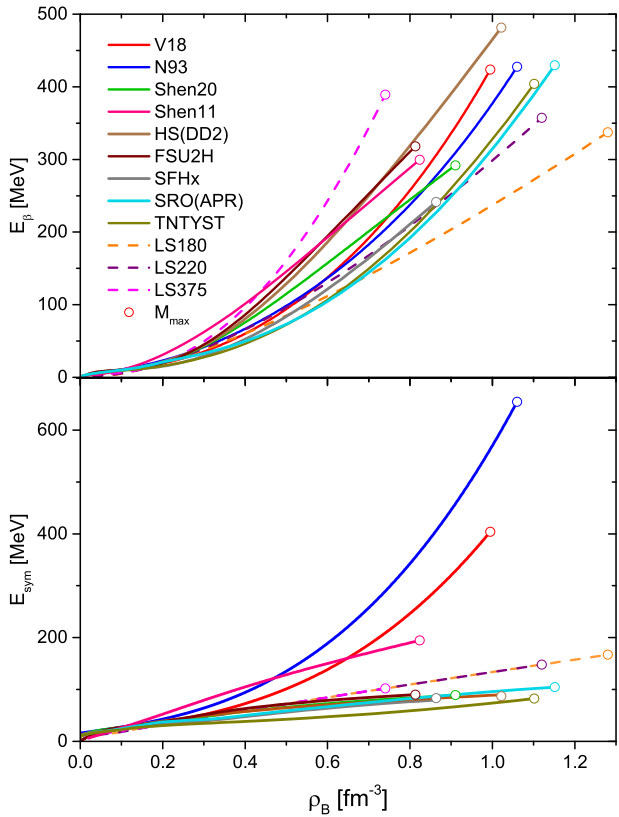}}
\vspace{-20mm}
\caption{
Energy per nucleon of $\beta$-stable matter $E_\beta$ (top panel)
and symmetry energy $\esym$ (bottom panel)
at $T=0$
for the different EOSs.
All curves end at the central density of their respective $\mmax$ configuration.
}
\label{f:e}
\end{figure}

The energy per nucleon of cold $\beta$-stable matter and the symmetry energy
are displayed in Fig.~\ref{f:e} for different EOSs.
One notes that at high density the symmetry energy is much larger
for the BHF EOSs V18 and N93 than for the other models,
which is an effect of particularly repulsive nuclear three-body forces
\cite{Li08a,Li08b}.

A particular feature of the set of LS EOSs can be appreciated here.
Those EOSs differ only in the compressibility of symmetric nuclear matter
as indicated by their name,
whereas other properties are unaffected by this choice,
in particular the symmetry energy as indicated in the lower panel,
and consequently proton fraction, partial particle densities,
and most important temperature effects.
This will be very helpful in the analysis of the theoretical results later on.
Regarding cold matter as shown in Fig.~\ref{f:e},
the three LS EOSs share the same symmetry energy,
but exhibit $\beta$-stable matter of rather different stiffness
and related maximum masses and corresponding central baryon densities.

At large enough density,
some of the non-relativistic EOSs become superluminal.
In our selection this happens,
before the central density of the NS maximum mass configuration is reached,
for the V18, N93, SRO(APR), TNTYST, and LS375.
However, in all cases the mass at the onset $c_s=1$ is close to $\mmax$ already:
This could be remedied by limiting $c_s<1$ by hand,
which yields very similar $\mmax$ as the original EOSs,
see, e.g., \cite{Sun21},
but we use the original unmodified EOSs in this work.


\begin{figure}[t]
\vspace{-9mm}
\centerline{\includegraphics[scale=0.38]{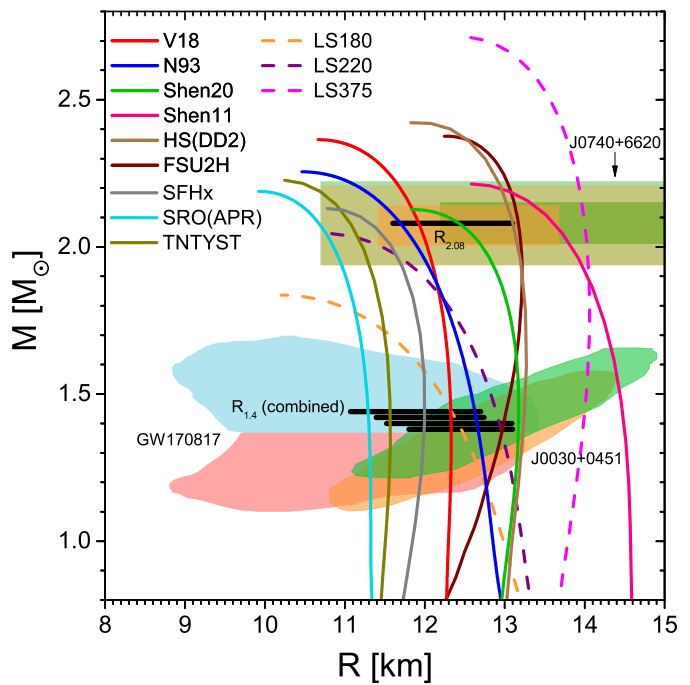}}
\vspace{-11mm}
\caption{
Mass-radius relations obtained with different EOSs.
The mass of the most heavy pulsar PSR J0740+6620 \cite{Cromartie20}
observed until now is also shown,
together with the constraints from the GW170817 event \cite{Abbott18}
and the mass-radius constraints
on the pulsars J0030+0451 and J0740+6620
of the NICER mission
\cite{Riley19,Miller19,Riley21,Miller21}.
The black bars indicate the limits on
$R_{2.08}$ and
$R_{1.4}$
obtained in the combined data analyses
of \cite{Miller21,Pang21,Raaijmakers21}.
}
\label{f:mr}
\end{figure}

\begin{table*}[t]
\renewcommand{\arraystretch}{1.15}
\caption{
\label{t:eos}
List of finite-temperature EOS models considered in this work.
For each model we mention properties of saturated symmetric matter
(density $\rho_0$; binding energy per nucleon $-E_0$;
compression modulus $K_0$; symmetry energy $S_0$;
slope of the symmetry energy $L$;
Landau effective mass of nucleons $m^*\!/m$; and also
neutron-proton effective mass splitting in neutron matter at $\rho_0$,
$\Delta m^*\!/m$).
Also given are key properties of nonrotating spherically-symmetric NSs
(maximum gravitational mass $\mmax$;
radius $R_{1.4}$ and tidal deformability $\Lambda_{1.4}$ of a $1.4\ms$ NS;
radius $R_{2.08}$ of a $2.08\ms$ NS).
Experimental nuclear parameters and observational constraints
from different sources are also listed for comparison.
See text for details.
}
\begin{ruledtabular}
\begin{tabular}{lcccccccccccc}
 Model  & Source/Ref. &  $\rho_0$ & $-E_0$ & $K_0$ & $S_0$ & $L$
        & $\mmax$ & $\Lambda_{1.4}$ & $R_{1.4}$ & $R_{2.08}$
        & $m^*\!/m$ & $\Delta m^*\!/m$ \\
	& &  $[\fm3]$ & [MeV] & [MeV] & [MeV] & [MeV] & $[\ms]$ &  & [km] & [km] &  &  \\
\hline
V18       & \cite{BHFWWW}  & 0.178  & 13.9 & 207 & 32.3 & 67 & 2.36 & 440 & 12.3 & 11.9 & 0.789 & 0.124 \\
N93       & \cite{BHFWWW}  & 0.185  & 16.1 & 229 & 36.5 & 77 & 2.25 & 470 & 12.7 & 11.7 & 0.739 & 0.139 \\
Shen11    & \cite{ShenWWW} & 0.145  & 16.3 & 281 & 36.9 &111 & 2.21 &1170 & 14.5 & 13.6 & 0.689 & 0.085 \\
Shen20    & \cite{ShenWWW} & 0.145  & 16.3 & 281 & 31.4 & 40 & 2.11 & 680 & 13.2 & 12.4 & 0.647 & 0.086 \\
SRO(APR)  & \cite{CompOSE} & 0.160  & 16.0 & 266 & 32.6 & 59 & 2.19 & 250 & 11.3 & 10.7 & 0.698 & 0.211 \\
TNTYST    & \cite{CompOSE} & 0.160  & 16.1 & 245 & 30.0 & 35 & 2.23 & 310 & 11.6 & 11.1 & -     & -     \\
SFH(SFHx)      & \cite{CompOSE} & 0.160  & 16.2 & 239 & 28.7 & 23 & 2.13 & 400 & 12.0 & 11.3 & 0.771 & 0.083 \\
HS(DD2)   & \cite{CompOSE} & 0.149  & 16.0 & 243 & 31.7 & 55 & 2.42 & 680 & 13.2 & 13.1 & 0.626 & 0.097 \\
FSU2H     & \cite{Tolos17} & 0.150  & 16.3 & 238 & 30.5 & 45 & 2.38 & 590 & 13.0 & 13.2 & 0.654 & 0.091 \\
\hline
 LS180    & \cite{LSWWW}   & 0.155  & 16.0 & 180 & 29.3 & 74 & 1.84 & 390 & 12.4 & -    & 1     & 0     \\
 LS220    & \cite{LSWWW}   & 0.155  & 16.0 & 220 & 29.3 & 74 & 2.04 & 540 & 12.9 & -    & 1     & 0     \\
 LS375    & \cite{LSWWW}   & 0.155  & 16.0 & 375 & 29.3 & 74 & 2.72 & 970 & 13.9 & 13.9 & 1     & 0     \\
\hline
 Exp.   &  & $\sim$ 0.14--0.17 & $\sim$ 15--16 & 220--260 & 28.5--34.9 & 30--87
        & $>2.14^{+0.10}_{-0.09}$
        & 70--580 & 11.1--12.7 & 11.6--13.1 & - & - \\
 Ref.   &  & \cite{Margueron18a} 
        & \cite{Margueron18a} 
        & \cite{Shlomo06,Piekarewicz10}
        & \cite{Li13,Oertel17} & \cite{Li13,Oertel17} & \cite{Cromartie20}
        & \cite{Abbott18} & \cite{Pang21} & \cite{Riley21} \\
\end{tabular}
\end{ruledtabular}
\end{table*}

The corresponding mass-radius relations of cold NSs are shown in Fig.~\ref{f:mr},
obtained in the standard way by solving the TOV equations for $\beta$-stable and
charge-neutral cold matter.
The figure also shows the latest observational contraints on mass and radius
from interpretation of the GW170817 event \cite{Abbott17,Abbott18}
and recent results of the NICER mission
\cite{Riley19,Miller19,Riley21,Miller21,Pang21,Raaijmakers21}.

The former allowed to constrain
the tidal deformability of the merging NSs,
and the related radius
$R_{1.36}=11.9\pm1.4\;$km \cite{Abbott18},
see also similar compatible constraints on masses and radii
derived in Refs.~\cite{%
Margalit17,  
Ruiz18,      
Most18,      
Rezzolla18,  
Shibata19,   
Most20}.     
Some theoretical analyses of this event
indicate also an upper limit on the maximum mass
of $\sim2.2-2.3\ms$ \cite{Shibata17,Margalit17,Rezzolla18,Shibata19}.
However, those are very model dependent,
in particular on the still to-be-determined temperature dependence of the EOS
\cite{Khadkikar_PRC_2021}.
It seems, however, that maximum masses
of static cold NSs
$\mtov\gtrsim2.4\ms$ can be excluded
by the simple fact that GW170817 did not leave a stable NS as a remnant
\cite{Figura21}.
Altogether, in this work we have chosen to consider as realistic EOSs
with $2.1<\mmax/\ms<2.4$
and those are plotted in all figures
and analyzed further in the following.

Regarding NICER,
so far results for the pulsars J0030+0451 \cite{Riley19,Miller19}
and J0740+6620 \cite{Riley21,Miller21,Pang21,Raaijmakers21}
are available and plotted in the figure.
The combined analysis of both pulsars together with GW170817
yields improved limits
on $R_{2.08}=12.35\pm0.75\;$km \cite{Miller21},
but in particular on the radius $R_{1.4}$,
namely
$12.45\pm0.65\;$km \cite{Miller21}, 
$11.94^{+0.76}_{-0.87}\;$km \cite{Pang21}, and
$12.33^{+0.76}_{-0.81}\;$km or
$12.18^{+0.56}_{-0.79}\;$km \cite{Raaijmakers21},
which are shown in Fig.~\ref{f:mr} as black horizontal bars.
The difference between these estimates reflects the model dependence
of the experimental analyses.
Note that taking seriously these constraints would further severly narrow
the set of compatible EOSs
and leave as realistic only the V18 and N93 EOS.
(There are of course several other compatible EOSs,
but without finite-temperature extensions that are analyzed here).
We summarize in Table~\ref{t:eos} the properties of the various EOSs
regarding these constraints.


\subsection{Hot neutron stars}

After selecting reasonable cold NS EOSs,
we now study the features of their extension to finite temperature.
As there are currently no directly applicable observational constraints,
those are rather varying purely theoretical predictions.

\begin{figure}[t]
\vspace{-14mm}
\centerline{\includegraphics[scale=0.35]{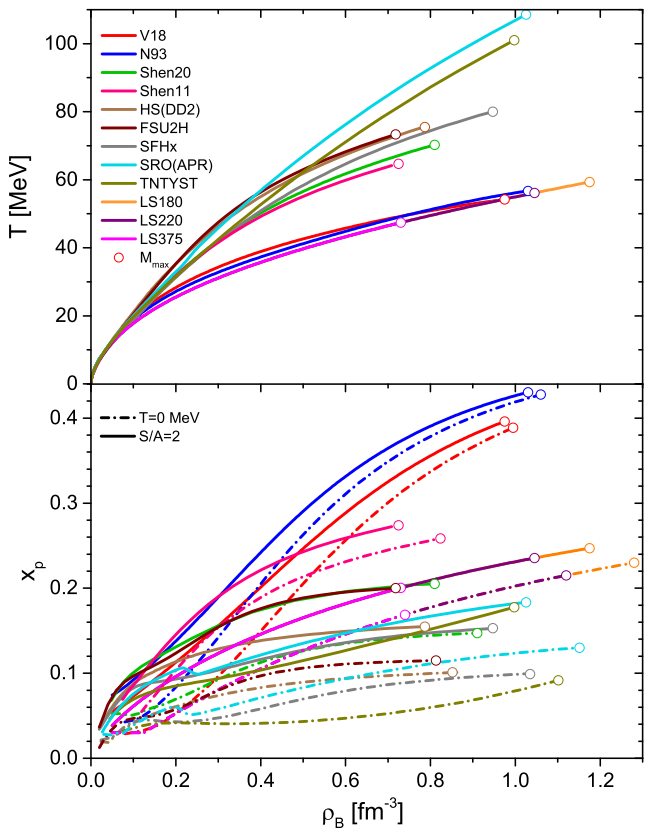}}
\vspace{-22mm}
\caption{
Upper panel:
The temperature profiles corresponding to $\beta$-stable matter with $S/A=2$.
Lower panel:
Proton fractions in $\beta$-stable matter
for $T=0$ (broken curves) and $S/A=2$ (solid curves).
Markers indicate the $\mmax$ configurations for each EOS.
}
\label{f:t}
\end{figure}

We begin with the temperature profile corresponding to the $S/A=2$ condition
for the various EOSs in the upper panel of Fig.~\ref{f:t}.

The fact that no other particle degree of freedom in addition to nucleons is
considered,
makes $T(\rho_B)$ decrease monotonically from the core to the outer layers.
At high densities a large dispersion is obtained among the predictions
of various EOS models.

In particular,
the highest temperatures at large density are reached by
the two variational EOSs, SRO(APR) and TNTYST,
for which values as high as 100 MeV are obtained.
The lowest temperatures are obtained for the two BHF EOSs and the three LS EOSs.
RMF models produce intermediate values.

At low density $\rho_B \lesssim 0.1\fm3$
the predictions of the different models converge.
This is the obvious consequence of good constraints on the low-density EOS
and relatively low values of temperature.
For low enough density
the temperature remains below the critical temperature $T_C$
of the liquid-gas phase transition,
which is a few MeV in asymmetric nuclear matter.
This means that, similar to what happens in cold NSs,
also in this case baryonic matter in the outer layers should be clusterized.
This is indeed the case of all general-purpose EOSs publicly available on
\cite{LSWWW,ShenWWW,CompOSE} that we use here.
The BHF and FSU2H EOSs have been complemented at low densities
in a thermodynamic consistent way
with the LS220 and HS(IUF) \cite{Hempel10} EOSs, respectively.
Clusterisation in the outer layers partially washes out the dependence on
effective interactions and introduces in exchange some dependence on
cluster modelling \cite{Raduta_2021}.

The bottom panel of Fig.~\ref{f:t} confronts the isospin asymmetry
(proton fraction)
of hot and cold $\beta$-stable matter.
First we notice that the cold proton fractions reflect the symmetry energy
shown in Fig.~\ref{f:e},
namely for the BHF models V18 and N93 they are substantially higher at large
density than for the other models.
Finite temperature increases the proton fractions,
in particular at low density,
where leptons become rather numerous
as a result of Fermi distributions at finite temperature.
Because of the charge-neutrality condition,
this increases the proton fraction
and thus the isospin symmetry of nuclear matter,
and this counteracts the stiffening of the EOS
due to the individual thermal pressures of the nucleons.
On the other hand,
the increase of the lepton densities with temperature
augments the thermal lepton pressure,
which in turn acts against the effect of increasing isospin symmetry.

At this point it is important to stress that,
for given particle degrees of freedom and
fixed values of density and entropy per baryon,
the temperature is correlated with the values of

\begin{figure}[t]
\vspace{-6mm}
\centerline{\includegraphics[scale=0.42]{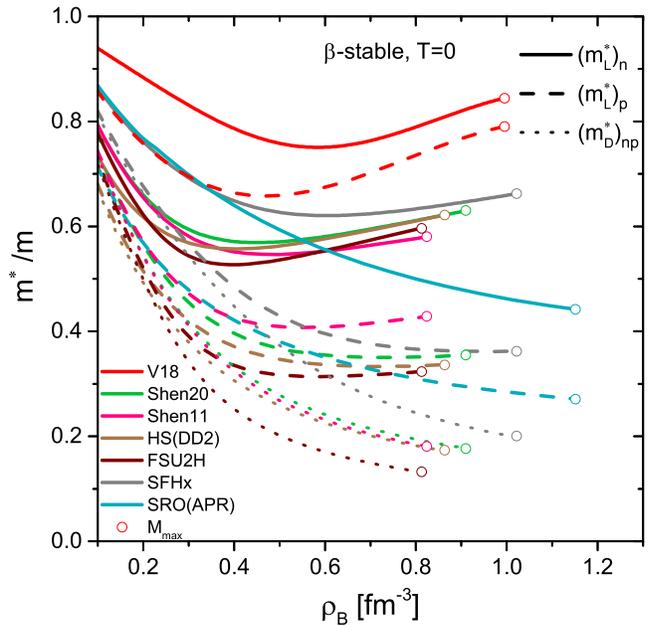}}
\vspace{-4mm}
\caption{
Neutron and proton effective masses in cold $\beta$-stable matter
as a function of the nucleon density for some EOSs considered in this work.
}
\label{f:meff}
\end{figure}

Landau (for non-relativistic models) or
Dirac (for relativistic models)
effective masses \cite{Raduta_2021}

and isospin asymmetry of the matter
\cite{Constantinou_PRC_2014,Constantinou_PRC_2015,Raduta20}.
Therefore the low (high) temperature values produced by the BHF (variational)
EOS models can be explained by the moderate (extreme) isospin asymmetries
that characterize these classes of models
(see the bottom panel of Fig.~\ref{f:t}), whereas
the low temperature values produced by the LS EOS models
are attributable to the high values of the Landau effective mass,
$m_L^*(\rho)/m=1$.
RMF models span a rather broad range of behaviors regarding
the density dependence of the symmetry energy,
which results in largely different values of the proton fraction,

and a certain dispersion in the density dependence of the Dirac effective mass,
see 
Fig.~\ref{f:meff}.
Comparison between Shen11, Shen20, and HS(DD2)
confirms the above commented correlation
between isospin asymmetry and temperature.

The relation between effective masses and temperature
can be understood
in the generally-used quasiparticle approximation for the entropy density,
\bal
 s =& - 2\sum_k \big( n(k) \ln n(k) + [1-n(k)] \ln [1-n(k)] \big) \:,
\eal
where $n(k)= \left\{ \exp{([e(k)-\tilde{\mu}]/T)} + 1 \right\}^{-1}$
stands for the Fermi-Dirac distribution,
{
$\tilde{\mu}$ represents the auxilliary chemical potentials,
and $e(k)$ corresponds to the kinetic term of the single-particle energies,
namely
$e(k)=k^2\!/2m_L^*$ in non-relativistic models and
$e(k)=\sqrt{m_D^{*2}+k^2}$ in relativistic models.
Therefore a reduced value of the Landau effective mass,
$m_L^*(\rho)/m \ll 1$,
will result in an increased steepness of $e(k)$ and,
for a fixed value of the temperature, a smaller entropy density.
For relativistic models the dependence of $e(k)$ on the Dirac effective mass
is much smaller, which prevents clear correlations to be established.

In Fig.~\ref{f:meff} we display the $n$ and $p$ Landau
effective masses as a function of the density for some of the EOSs.
(Effective masses of N93 and TNTYST are not available;
see Fig.~9 in \cite{Kanzawa07} for the latter model.)
In BHF EOSs the effective Landau mass first decreases with increasing density,
and then it increases because of the enhanced repulsion,
mediated by three-body forces \cite{Baldo14}.
Qualitatively the same pattern characterizes RMF models,
but for a completely different reason,
namely the dominance of momentum over mass,
which is a genuine relativistic effect.
Finally in APR the effective Landau mass decreases with density.
The same is true for Skyrme-like interactions,
although the density dependence is different \cite{Raduta_2021}.

Indeed the theoretical expectation is confirmed,
namely, consistent with the temperature profiles,
the BHF (variational) EOSs feature the largest (smallest) effective masses
at high density.
For completeness the figure shows also
the $n$ and $p$ Dirac effective masses
for the RMF models.
Due to the absence of scalar-isovector mesons in all considered models,
both quantities coincide,
and strongly decrease with density due to enhanced interactions.
}

\begin{figure*}[t]
\vspace{-6mm}
\centerline{\includegraphics[scale=0.30]{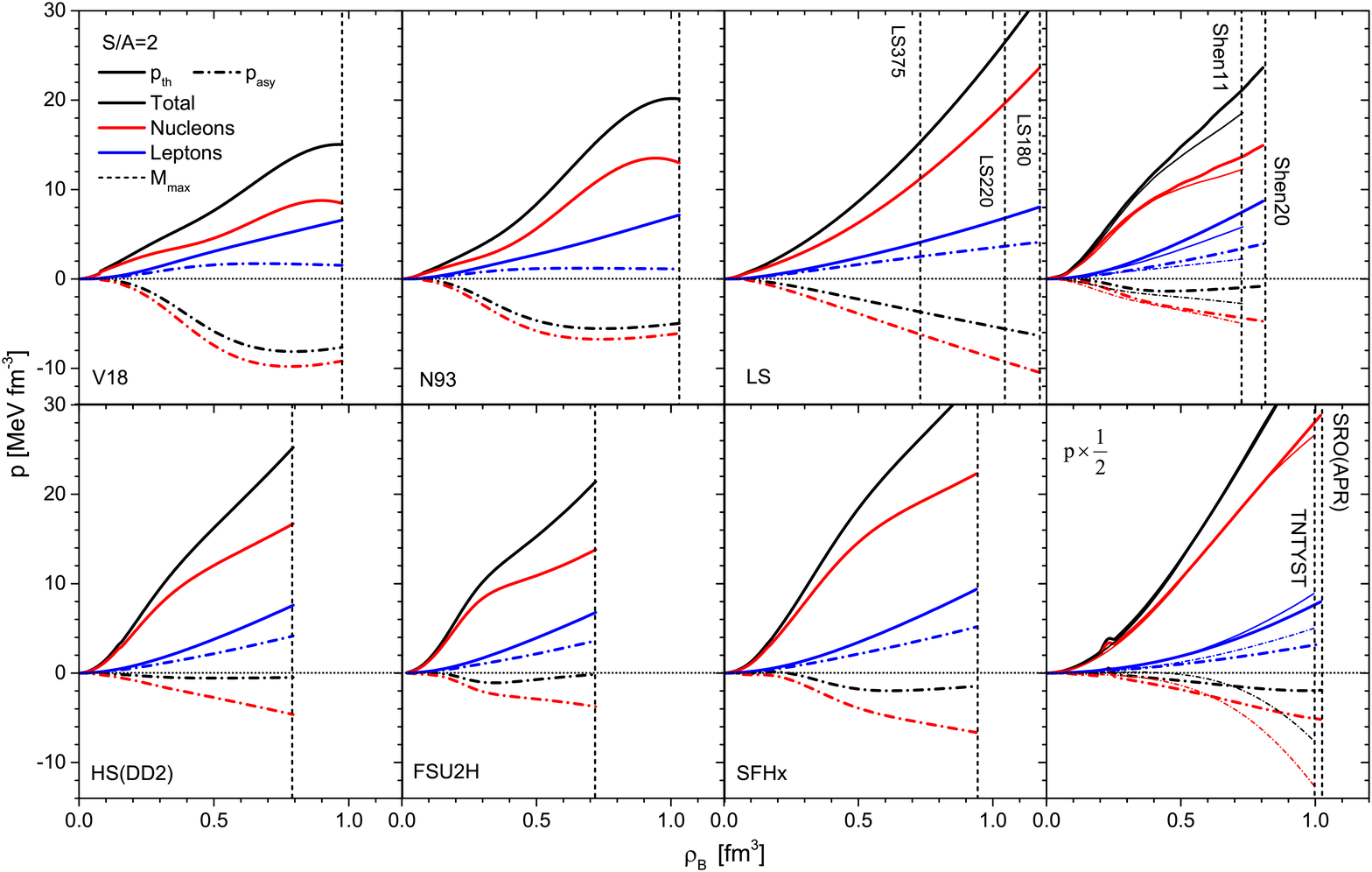}}
\vspace{-17mm}
\caption{
Nucleon and lepton contributions to the
thermal pressure of $\beta$-stable matter,
Eq.~(\ref{e:pth}),
at $S/A=2$ 
for the different EOSs.
The dot-dashed curves show the pressure caused by composition change only,
Eq.~(\ref{e:pasy}).
The vertical lines indicate the central density of the $\mmax$ configurations.
The SRO(APR) and TNTYST results in the last panel are scaled down
by a factor of 2.}
\label{f:p}
\end{figure*}

\begin{figure}[t]
\vspace{-3mm}
\centerline{\includegraphics[scale=0.43]{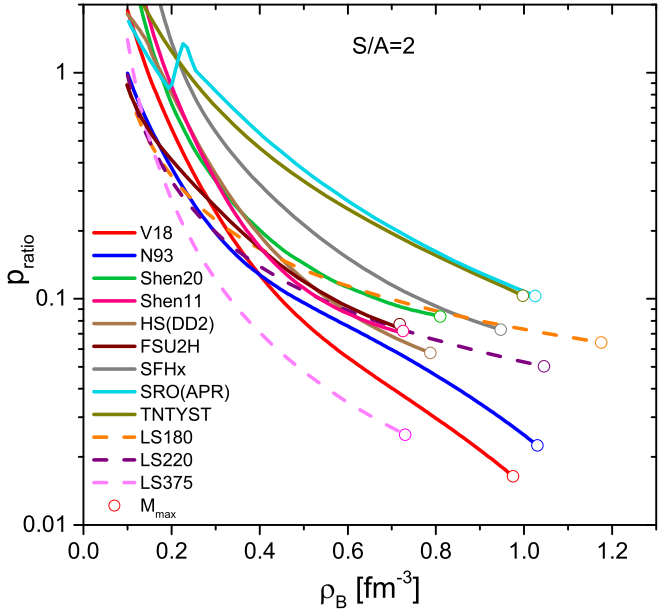}}
\vspace{-5mm}
\caption{
Ratio of thermal pressure to cold pressure of $\beta$-stable matter,
Eq.~(\ref{e:pr}),
as a function of density
for different EOSs.
All curves end at the central density of their respective $\mmax$ configuration.
}
\label{f:pr}
\end{figure}

\begin{figure*}[t]
\vspace{-12mm}
\centerline{\hspace{0mm}\includegraphics[scale=0.3]{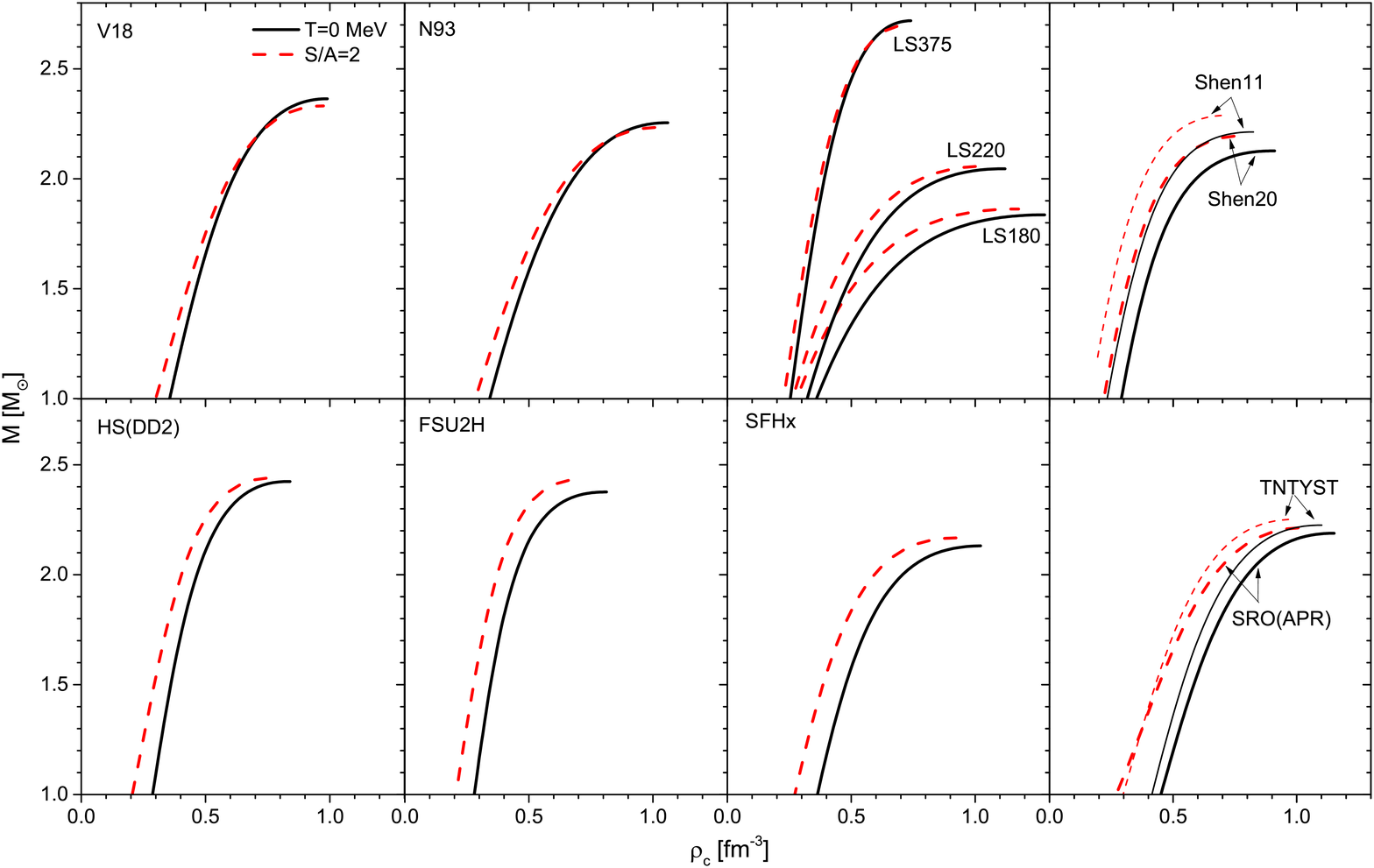}}
\vspace{-13mm}
\caption{
NS gravitational mass vs central density
for cold and hot stars and the different EOSs.}
\label{f:m}
\end{figure*}

The interplay between thermal effects and modification of chemical composition
is illustrated in Fig.~\ref{f:p},
where we display the different contributions to the thermal pressure
of $\beta$-stable matter (solid curves),
\be
 \pt(\rho,T) \equiv p(\rho,x_T,T) - p(\rho,x_0,0) \:,
\label{e:pth}
\ee
where $x_T$ and $x_0$ are the respective proton fractions of
hot ($S/A=2$) and cold matter,
as shown in Fig.~\ref{f:t}.
We do not illustrate here the contribution of the hot photon gas,
as for the thermodynamic conditions under consideration it is negligible:
at $T=50\;$MeV the associated thermal energy density and pressure are
$\eps/3 = p = \pi^2 T^4/45$      
and thus $\sim 0.2\mfm$.
We nevertheless mention that these are included in our calculations
as well as in the publicly available general-purpose EOS tables.

It is useful to also analyze the ``asymmetry thermal pressure"
for nucleons and leptons separately,
namely (dot-dashed curves)
\be
 p_\text{asy}(\rho,T) \equiv  p(\rho,x_T,0) - p(\rho,x_0,0) \:,
\label{e:pasy}
\ee
i.e., the contribution to the thermal pressure caused only
by the change of the composition, while keeping the matter cold.
The thermal pressure can then be expressed as
\be
 \pt(\rho,T) = p_\text{asy}(\rho,T) + [p(\rho,x_T,T) - p(\rho,x_T,0)] \:,
\ee
i.e., adding the thermal partial pressures at fixed composition $x_T$.
One can clearly see that
the nucleonic and total $p_\text{asy}$ is negative for all EOSs,
and thus the change of composition of hot $\beta$-stable matter
(increased isospin symmetry, i.e., proton fraction)
reduces the pressure of the matter, even including the leptons.
Adding the individual thermal pressures of nucleons and leptons
at fixed partial densities
then provides a total thermal pressure that is reduced
compared to the one neglecting composition change.
For the BHF EOSs with large proton fractions,
the effect of pressure reduction by
composition change is particularly strong,
leading to a relatively small total thermal pressure.
On the other hand,
the variational EOSs exhibit by far the largest thermal pressure,
mainly because of their high temperature for the fixed-entropy profile.

The thermal pressure is an important quantity,
because it potentially increases the stability of the star against collapse,
i.e., it might increase the maximum gravitational mass of the hot NS
\cite{Kaplan14,Schneider20,Raithel19,Raithel19_Err}.
However, it competes with the Fermi pressure of the cold matter,
and therefore a reasonable quantity to analyze is the relative thermal pressure
\cite{Raithel21}
\be
 p_\text{ratio} \equiv \frac{\pt}{p_0}(\rho,T)
 = \frac{p(\rho,x_T,T) - p(\rho,x_0,0)}{p(\rho,x_0,0)} \:,
\label{e:pr}
\ee
which is displayed in Fig.~\ref{f:pr} for the various EOSs.
With the exception of SRO(APR) at
$\rho \approx 0.2 - 0.3\fm3$,
where the transition to a pion condensate produces an irregularity,
$p_\text{ratio}$ naturally decreases with increasing density
and reaches a few percent at the maximum-mass configurations.
But these minimum values are rather varied for the different EOSs:
The V18, N93, and LS375 ratios are below 3 percent,
while the others are up to 10 percent
for the SRO(APR) and TNTYST.
The result for the set of LS EOSs is particulary illuminating,
as their thermal pressure is identical for all three models,
see Fig.~\ref{f:p},
and the different ratios are caused solely by different Fermi pressures
of cold matter,
related to the different incompressibility values.

After this discussion of the features of the temperature-dependent
$\beta$-stable EOSs,
we now focus on the consequences for stellar structure.
Fig.~\ref{f:m} shows the gravitational mass - central density relations
for cold and hot NSs and the different EOSs.
In general the effect of finite temperature is small,
given the typical Fermi energies involved.
In particular the change of the maximum masses is only a few percent
and can be both positive and negative.

\begin{figure}[t]
\vspace{-9mm}
\centerline{\hspace{9mm}\includegraphics[scale=0.38,clip]{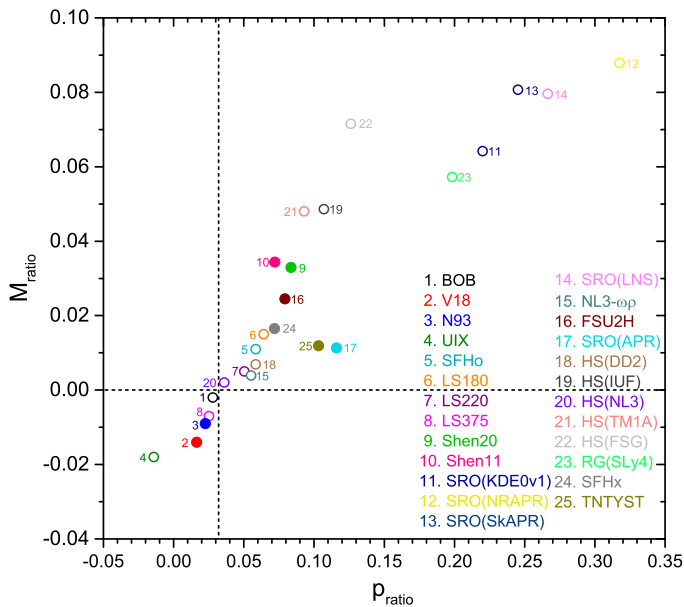}}
\vspace{-9mm}
\caption{
Correlation between relative change of the maximum gravitational mass,
Eq.~(\ref{e:mr}),
and the pressure ratio at the center of the star,
Eq.~(\ref{e:pr}),
for different EOSs.
Full symbols indicate EOSs respecting the constraint
$2.1<\mmax/\ms<2.4$.
}
\label{f:dm}
\end{figure}

The relative change of the gravitational maximum mass
\be
 M_\text{ratio} \equiv
 \frac{\mmax^\text{hot}-\mmax^\text{cold}}{\mmax^\text{cold}}
\label{e:mr}
\ee
is plotted in Fig.~\ref{f:dm} against the thermal pressure ratio
of Eq.~(\ref{e:pr}), shown in Fig.~\ref{f:pr},
taken at the central density of the $\mmax^\text{hot}$ star.
One observes a clear correlation, namely EOSs with
$p_\text{ratio}\lesssim0.03$ lead to a decrease of $\mmax$
and vice versa.

In this plot we also show the results
(open markers)
obtained for several other EOSs
that do not fulfill our selection critera on NS mass and radius
(in most cases $\mmax$ is too small).
Those EOSs are
BOB and UIX \cite{Lu19},
RG(SLy4) \cite{Raduta_NPA_2019,SLy4},
SRO(KDE0v1) \cite{SRO_PRC_2017,KDE0v1},
SRO(NRAPR) \cite{SRO_PRC_2017,Schneider19},
SRO(SkAPR) \cite{SRO_PRC_2017},
SRO(LNS) \cite{SRO_PRC_2017,LNS},
NL3-$\omega\rho$ \cite{Pais16,Horowitz01},
HS(IUF) \cite{Hempel10,IUF},
HS(NL3) \cite{Hempel10,Lalazissis_PRC_1997},
HS(FSG) \cite{Hempel10,Todd-Rutel_PRL_2005},
HS(TM1A) \cite{Hempel10,TM1A}.
Except BOB, UIX, and NL3-$\omega\rho$,
all these EOSs are publicly available on \textsc{CompOSE} \cite{Typel_2013}.
The correlation between $M_\text{ratio}$ and $p_\text{ratio}$ clearly persists
for all of them,
with many of the EOSs exhibiting an unnaturaly high $M_\text{ratio}$
due to the fact that their $\mmax$ and the correlated cold Fermi pressures
are too low and therefore $p_\text{ratio}$ too high.
Heating obviously produces larger changes
to a low-mass star
than to a high-mass star
for fixed thermal conditions.
For the subset of realistic EOSs
(plotted using full markers),
the relative increase of $\mmax$ is limited to less than 4 percent.

\medskip\medskip
\section{Summary}
\label{s:end}

We examined in this work several nuclear EOSs
extended to finite temperature
and suitable to model NSs
under different physical conditions.
We focussed this study on $\beta$-stable isentropic matter with $S/A=2$,
which involves temperatures typical for supernovae, proto NSs, and NS mergers,
for example.

We first singled out the cold EOSs compatible with the most recent
astrophysical constraints from GW170817 and the NICER mission.
Only a few EOSs satisfy simultaneously the lower limits on the maximum mass
and the current constraints on the radii $R_{1.4}$ and $R_{2.08}$.
We then studied the finite-temperature extensions of those EOSs.

We concluded that the BHF EOSs V18 and N93 with large symmetry energies
feature particularly low ratios of thermal pressure to Fermi pressure,
and consequently predict no increase of stellar stability due to finite
temperature, in contrast to most other EOSs, apart from the LS375.
This is due to an important reduction of the thermal pressure
by the increase of isospin symmetry in hot nuclear matter in this case,
which is a consequence of $\beta$-stability
and as such relevant for physical systems
that evolve slow enough to maintain this condition.
On the contrary,
temperature effects are large for the variational TNTYST and SRO(APR) EOSs,
due to their small effective masses and symmetry energies.
However, predicted NS radii are rather (too) small in this case.

A detailed study of the family of LS EOSs revealed clearly that the
properties of the finite-temperature extension of an EOS alone
(which is the same for all three LS EOSs)
cannot be correlated with the impact on stellar stability,
but that the cold EOS is of equal importance for this feature.

All EOSs we studied
(also those not fulfilling current mass and radius constraints)
exhibit a clear correlation between
the ratio of thermal and Fermi pressures at the NS maximum central density
and the relative change of the maximum gravitational mass:
The mass increases only for EOSs with a pressure ratio larger
than about 3 percent.
For the subset of realistic EOSs,
the relative change of the maximum gravitational mass
varies between about $-2$ and $+4$ percent,
depending on details of the EOS and its extension to finite temperature.
This is not a big effect, and it will require serious efforts
to obtain useful constraints on this feature
from future astrophysical observations.

The physical conditions we examined in this study were rather idealized:
$\beta$-stable and isentropic matter.
This allowed to isolate clearly the consequences of finite temperature,
but in particular the effect of neutrino trapping was completely
disregarded in this schematic investigation.
It is well known that
the neutrino contributions to thermal energy density and pressure
might also cause a substantial increase of the maximum masses
\cite{Nicotra06,Li10,Burgio11,Paschalidis12,Kaplan14,Marques17,Lalit19}.
But for a consistent analysis detailed microscopic simulations
suitable for the physical system under investigation
(proto NS, merger, ...) are required.
{
First results indicate that $\beta$-stability is violated in
numerical simulations of the GW170817 merger event \cite{Figura20,Figura21}.

A further important point to consider is the hypothetical appearance of
strange baryonic matter in the NS core.
A few finite-temperature EOSs including hyperons
are available \cite{Ishizuka08,Burgio2011,Oertel_EPJA_2016},
but not all comply with the $\approx 2\ms$ lower limit on the
maximum gravitational mass of cold non-rotating NSs.
In any case, from many past studies it is known that the presence
of hyperons increases the maximum mass of hot NSs,
allowing a possible `delayed collapse' phenomenon during the proto NS cooldown,
see, e.g., \cite{Prakash97,Burgio2011,Raduta20} and references therein.
}

With the expected rapid progress in astrophysical observations
due to new techniques and instruments,
the confrontation of numerical simulations with new data
will allow to narrow down the features of the EOS of cold matter
and its extension to finite temperature more and more
in the close future.

\section*{Acknowledgments}

We acknowledge
financial support from the China Scholarship Council (No.~201706410092)
and COST Action CA16214 ``PHAROS''.
A.R.R.~acknowledges support from a grant of the Ministry of Research,
Innovation and Digitization, CNCS/CCCDI - UEFISCDI,
Project No. PN-III-P4-ID-PCE-2020-0293, within PNCDI III.
G.F.B.~and H.-J.S.~acknowledge financial contribution from the agreement
ASI-INAF n.2017-14-H.O.

\newcommand{\physrep}{Phys.~Rep.}
\newcommand{\nphysa}{Nucl.~Phys.~A}
\newcommand{\npa}{Nucl.~Phys.~A}
\newcommand{\aap}{A\&A}
\newcommand{\mnras}{MNRAS}
\newcommand{\epja}{EPJA}
\newcommand{\araa}{Annu. Rev. Astron.~Astrophys.}
\newcommand{\apjs}{ApJS}
\newcommand{\apjl}{Astrophys.~J.~Lett.}
\newcommand{\rpp}{Rep. Prog. Phys.}
\newcommand{\ppnp}{Prog. Part. Nucl. Phys.}
\newcommand{\plb}{Phys. Lett. B}
\bibliographystyle{apsrev4-1}
\bibliography{hoteos}

\begin{thebibliography}{88}%
\makeatletter
\providecommand \@ifxundefined [1]{%
 \@ifx{#1\undefined}
}%
\providecommand \@ifnum [1]{%
 \ifnum #1\expandafter \@firstoftwo
 \else \expandafter \@secondoftwo
 \fi
}%
\providecommand \@ifx [1]{%
 \ifx #1\expandafter \@firstoftwo
 \else \expandafter \@secondoftwo
 \fi
}%
\providecommand \natexlab [1]{#1}%
\providecommand \enquote  [1]{``#1''}%
\providecommand \bibnamefont  [1]{#1}%
\providecommand \bibfnamefont [1]{#1}%
\providecommand \citenamefont [1]{#1}%
\providecommand \href@noop [0]{\@secondoftwo}%
\providecommand \href [0]{\begingroup \@sanitize@url \@href}%
\providecommand \@href[1]{\@@startlink{#1}\@@href}%
\providecommand \@@href[1]{\endgroup#1\@@endlink}%
\providecommand \@sanitize@url [0]{\catcode `\\12\catcode `\$12\catcode
  `\&12\catcode `\#12\catcode `\^12\catcode `\_12\catcode `\%12\relax}%
\providecommand \@@startlink[1]{}%
\providecommand \@@endlink[0]{}%
\providecommand \url  [0]{\begingroup\@sanitize@url \@url }%
\providecommand \@url [1]{\endgroup\@href {#1}{\urlprefix }}%
\providecommand \urlprefix  [0]{URL }%
\providecommand \Eprint [0]{\href }%
\providecommand \doibase [0]{http://dx.doi.org/}%
\providecommand \selectlanguage [0]{\@gobble}%
\providecommand \bibinfo  [0]{\@secondoftwo}%
\providecommand \bibfield  [0]{\@secondoftwo}%
\providecommand \translation [1]{[#1]}%
\providecommand \BibitemOpen [0]{}%
\providecommand \bibitemStop [0]{}%
\providecommand \bibitemNoStop [0]{.\EOS\space}%
\providecommand \EOS [0]{\spacefactor3000\relax}%
\providecommand \BibitemShut  [1]{\csname bibitem#1\endcsname}%
\let\auto@bib@innerbib\@empty
\bibitem [{\citenamefont {{B. P. Abbott {\it et al.}}}(2017)}]{Abbott17}%
  \BibitemOpen
  \bibfield  {author} {\bibinfo {author} {\bibnamefont {{B. P. Abbott {\it et
  al.}}}},\ }\href@noop {} {\bibfield  {journal} {\bibinfo  {journal} {Phys.
  Rev. Lett.}\ }\textbf {\bibinfo {volume} {119}},\ \bibinfo {pages} {161101}
  (\bibinfo {year} {2017})}\BibitemShut {NoStop}%
\bibitem [{\citenamefont {Abbott}\ \emph {et~al.}(2018)\citenamefont {Abbott}
  \emph {et~al.}}]{Abbott18}%
  \BibitemOpen
  \bibfield  {author} {\bibinfo {author} {\bibfnamefont {B.~P.}\ \bibnamefont
  {Abbott}} \emph {et~al.},\ }\href@noop {} {\bibfield  {journal} {\bibinfo
  {journal} {\prl}\ }\textbf {\bibinfo {volume} {121}},\ \bibinfo {eid}
  {161101} (\bibinfo {year} {2018})}\BibitemShut {NoStop}%
\bibitem [{\citenamefont {{T. E. Riley {\it et al.}}}(2019)}]{Riley19}%
  \BibitemOpen
  \bibfield  {author} {\bibinfo {author} {\bibnamefont {{T. E. Riley {\it et
  al.}}}},\ }\href@noop {} {\bibfield  {journal} {\bibinfo  {journal}
  {Astrophys. J. Lett.}\ }\textbf {\bibinfo {volume} {887}},\ \bibinfo {pages}
  {L21} (\bibinfo {year} {2019})}\BibitemShut {NoStop}%
\bibitem [{\citenamefont {{M. C. Miller {\it et al.}}}(2019)}]{Miller19}%
  \BibitemOpen
  \bibfield  {author} {\bibinfo {author} {\bibnamefont {{M. C. Miller {\it et
  al.}}}},\ }\href@noop {} {\bibfield  {journal} {\bibinfo  {journal}
  {Astrophys. J. Lett.}\ }\textbf {\bibinfo {volume} {887}},\ \bibinfo {pages}
  {L24} (\bibinfo {year} {2019})}\BibitemShut {NoStop}%
\bibitem [{\citenamefont {Riley}\ \emph {et~al.}(2021)\citenamefont {Riley}
  \emph {et~al.}}]{Riley21}%
  \BibitemOpen
  \bibfield  {author} {\bibinfo {author} {\bibfnamefont {T.~E.}\ \bibnamefont
  {Riley}} \emph {et~al.},\ }\href {\doibase 10.3847/2041-8213/ac0a81}
  {\bibfield  {journal} {\bibinfo  {journal} {Astrophys. J. Lett.}\ }\textbf
  {\bibinfo {volume} {918}},\ \bibinfo {pages} {L27} (\bibinfo {year}
  {2021})}\BibitemShut {NoStop}%
\bibitem [{\citenamefont {Miller}\ \emph {et~al.}(2021)\citenamefont {Miller}
  \emph {et~al.}}]{Miller21}%
  \BibitemOpen
  \bibfield  {author} {\bibinfo {author} {\bibfnamefont {M.~C.}\ \bibnamefont
  {Miller}} \emph {et~al.},\ }\href {\doibase 10.3847/2041-8213/ac089b}
  {\bibfield  {journal} {\bibinfo  {journal} {Astrophys. J. Lett.}\ }\textbf
  {\bibinfo {volume} {918}},\ \bibinfo {pages} {L28} (\bibinfo {year}
  {2021})}\BibitemShut {NoStop}%
\bibitem [{\citenamefont {{Akmal}}\ \emph {et~al.}(1998)\citenamefont
  {{Akmal}}, \citenamefont {{Pandharipande}},\ and\ \citenamefont
  {{Ravenhall}}}]{Akmal98}%
  \BibitemOpen
  \bibfield  {author} {\bibinfo {author} {\bibfnamefont {A.}~\bibnamefont
  {{Akmal}}}, \bibinfo {author} {\bibfnamefont {V.~R.}\ \bibnamefont
  {{Pandharipande}}}, \ and\ \bibinfo {author} {\bibfnamefont {D.~G.}\
  \bibnamefont {{Ravenhall}}},\ }\href {\doibase 10.1103/PhysRevC.58.1804}
  {\bibfield  {journal} {\bibinfo  {journal} {\prc}\ }\textbf {\bibinfo
  {volume} {58}},\ \bibinfo {pages} {1804} (\bibinfo {year}
  {1998})}\BibitemShut {NoStop}%
\bibitem [{\citenamefont {{M. Maggiore {\it et al.}}}(2020)}]{Maggiore20}%
  \BibitemOpen
  \bibfield  {author} {\bibinfo {author} {\bibnamefont {{M. Maggiore {\it et
  al.}}}},\ }\href@noop {} {\bibfield  {journal} {\bibinfo  {journal} {JCAP}\
  }\textbf {\bibinfo {volume} {03}},\ \bibinfo {pages} {050} (\bibinfo {year}
  {2020})}\BibitemShut {NoStop}%
\bibitem [{\citenamefont {{Oertel}}\ \emph {et~al.}(2017)\citenamefont
  {{Oertel}}, \citenamefont {{Hempel}}, \citenamefont {{Kl\"ahn}},\ and\
  \citenamefont {{Typel}}}]{Oertel17}%
  \BibitemOpen
  \bibfield  {author} {\bibinfo {author} {\bibfnamefont {M.}~\bibnamefont
  {{Oertel}}}, \bibinfo {author} {\bibfnamefont {M.}~\bibnamefont {{Hempel}}},
  \bibinfo {author} {\bibfnamefont {T.}~\bibnamefont {{Kl\"ahn}}}, \ and\
  \bibinfo {author} {\bibfnamefont {S.}~\bibnamefont {{Typel}}},\ }\href
  {\doibase 10.1103/RevModPhys.89.015007} {\bibfield  {journal} {\bibinfo
  {journal} {Rev. Mod. Phys.}\ }\textbf {\bibinfo {volume} {89}},\ \bibinfo
  {eid} {015007} (\bibinfo {year} {2017})}\BibitemShut {NoStop}%
\bibitem [{\citenamefont {Burgio}\ \emph {et~al.}(2021)\citenamefont {Burgio},
  \citenamefont {Schulze}, \citenamefont {Vida{\~n}a},\ and\ \citenamefont
  {Wei}}]{Burgio21}%
  \BibitemOpen
  \bibfield  {author} {\bibinfo {author} {\bibfnamefont {G.}~\bibnamefont
  {Burgio}}, \bibinfo {author} {\bibfnamefont {H.-J.}\ \bibnamefont {Schulze}},
  \bibinfo {author} {\bibfnamefont {I.}~\bibnamefont {Vida{\~n}a}}, \ and\
  \bibinfo {author} {\bibfnamefont {J.-B.}\ \bibnamefont {Wei}},\ }\href@noop
  {} {\bibfield  {journal} {\bibinfo  {journal} {\ppnp}\ }\textbf {\bibinfo
  {volume} {120}},\ \bibinfo {pages} {103879} (\bibinfo {year}
  {2021})}\BibitemShut {NoStop}%
\bibitem [{\citenamefont {{Togashi}}\ \emph {et~al.}(2017)\citenamefont
  {{Togashi}}, \citenamefont {{Nakazato}}, \citenamefont {{Takehara}},
  \citenamefont {{Yamamuro}}, \citenamefont {{Suzuki}},\ and\ \citenamefont
  {{Takano}}}]{Togashi17}%
  \BibitemOpen
  \bibfield  {author} {\bibinfo {author} {\bibfnamefont {H.}~\bibnamefont
  {{Togashi}}}, \bibinfo {author} {\bibfnamefont {K.}~\bibnamefont
  {{Nakazato}}}, \bibinfo {author} {\bibfnamefont {Y.}~\bibnamefont
  {{Takehara}}}, \bibinfo {author} {\bibfnamefont {S.}~\bibnamefont
  {{Yamamuro}}}, \bibinfo {author} {\bibfnamefont {H.}~\bibnamefont
  {{Suzuki}}}, \ and\ \bibinfo {author} {\bibfnamefont {M.}~\bibnamefont
  {{Takano}}},\ }\href {\doibase 10.1016/j.nuclphysa.2017.02.010} {\bibfield
  {journal} {\bibinfo  {journal} {\nphysa}\ }\textbf {\bibinfo {volume}
  {961}},\ \bibinfo {pages} {78} (\bibinfo {year} {2017})}\BibitemShut
  {NoStop}%
\bibitem [{\citenamefont {{Schneider}}\ \emph {et~al.}(2019)\citenamefont
  {{Schneider}}, \citenamefont {{Constantinou}}, \citenamefont {{Muccioli}},\
  and\ \citenamefont {{Prakash}}}]{Schneider19}%
  \BibitemOpen
  \bibfield  {author} {\bibinfo {author} {\bibfnamefont {A.~S.}\ \bibnamefont
  {{Schneider}}}, \bibinfo {author} {\bibfnamefont {C.}~\bibnamefont
  {{Constantinou}}}, \bibinfo {author} {\bibfnamefont {B.}~\bibnamefont
  {{Muccioli}}}, \ and\ \bibinfo {author} {\bibfnamefont {M.}~\bibnamefont
  {{Prakash}}},\ }\href {\doibase 10.1103/PhysRevC.100.025803} {\bibfield
  {journal} {\bibinfo  {journal} {\prc}\ }\textbf {\bibinfo {volume} {100}},\
  \bibinfo {eid} {025803} (\bibinfo {year} {2019})}\BibitemShut {NoStop}%
\bibitem [{\citenamefont {Wiringa}\ \emph {et~al.}(1995)\citenamefont
  {Wiringa}, \citenamefont {Stoks},\ and\ \citenamefont
  {Schiavilla}}]{ArgonneV18}%
  \BibitemOpen
  \bibfield  {author} {\bibinfo {author} {\bibfnamefont {R.~B.}\ \bibnamefont
  {Wiringa}}, \bibinfo {author} {\bibfnamefont {V.~G.~J.}\ \bibnamefont
  {Stoks}}, \ and\ \bibinfo {author} {\bibfnamefont {R.}~\bibnamefont
  {Schiavilla}},\ }\href {\doibase 10.1103/PhysRevC.51.38} {\bibfield
  {journal} {\bibinfo  {journal} {Phys. Rev. C}\ }\textbf {\bibinfo {volume}
  {51}},\ \bibinfo {pages} {38} (\bibinfo {year} {1995})}\BibitemShut {NoStop}%
\bibitem [{\citenamefont {Carlson}\ \emph {et~al.}(1983)\citenamefont
  {Carlson}, \citenamefont {Pandharipande},\ and\ \citenamefont
  {Wiringa}}]{Carlson_NPA_1983}%
  \BibitemOpen
  \bibfield  {author} {\bibinfo {author} {\bibfnamefont {J.}~\bibnamefont
  {Carlson}}, \bibinfo {author} {\bibfnamefont {V.~R.}\ \bibnamefont
  {Pandharipande}}, \ and\ \bibinfo {author} {\bibfnamefont {R.~B.}\
  \bibnamefont {Wiringa}},\ }\href {\doibase 10.1016/0375-9474(83)90336-6}
  {\bibfield  {journal} {\bibinfo  {journal} {Nucl. Phys. A}\ }\textbf
  {\bibinfo {volume} {401}},\ \bibinfo {pages} {59} (\bibinfo {year}
  {1983})}\BibitemShut {NoStop}%
\bibitem [{\citenamefont {Pudliner}\ \emph {et~al.}(1995)\citenamefont
  {Pudliner}, \citenamefont {Pandharipande}, \citenamefont {Carlson},\ and\
  \citenamefont {Wiringa}}]{Pudliner_PRL_1995}%
  \BibitemOpen
  \bibfield  {author} {\bibinfo {author} {\bibfnamefont {B.~S.}\ \bibnamefont
  {Pudliner}}, \bibinfo {author} {\bibfnamefont {V.~R.}\ \bibnamefont
  {Pandharipande}}, \bibinfo {author} {\bibfnamefont {J.}~\bibnamefont
  {Carlson}}, \ and\ \bibinfo {author} {\bibfnamefont {R.~B.}\ \bibnamefont
  {Wiringa}},\ }\href {\doibase 10.1103/PhysRevLett.74.4396} {\bibfield
  {journal} {\bibinfo  {journal} {Phys. Rev. Lett.}\ }\textbf {\bibinfo
  {volume} {74}},\ \bibinfo {pages} {4396} (\bibinfo {year}
  {1995})}\BibitemShut {NoStop}%
\bibitem [{\citenamefont {{Li}}\ \emph {et~al.}(2008)\citenamefont {{Li}},
  \citenamefont {{Lombardo}}, \citenamefont {{Schulze}},\ and\ \citenamefont
  {{Zuo}}}]{Li08a}%
  \BibitemOpen
  \bibfield  {author} {\bibinfo {author} {\bibfnamefont {Z.~H.}\ \bibnamefont
  {{Li}}}, \bibinfo {author} {\bibfnamefont {U.}~\bibnamefont {{Lombardo}}},
  \bibinfo {author} {\bibfnamefont {H.-J.}\ \bibnamefont {{Schulze}}}, \ and\
  \bibinfo {author} {\bibfnamefont {W.}~\bibnamefont {{Zuo}}},\ }\href
  {\doibase 10.1103/PhysRevC.77.034316} {\bibfield  {journal} {\bibinfo
  {journal} {\prc}\ }\textbf {\bibinfo {volume} {77}},\ \bibinfo {eid} {034316}
  (\bibinfo {year} {2008})}\BibitemShut {NoStop}%
\bibitem [{\citenamefont {{Li}}\ and\ \citenamefont {{Schulze}}(2008)}]{Li08b}%
  \BibitemOpen
  \bibfield  {author} {\bibinfo {author} {\bibfnamefont {Z.~H.}\ \bibnamefont
  {{Li}}}\ and\ \bibinfo {author} {\bibfnamefont {H.-J.}\ \bibnamefont
  {{Schulze}}},\ }\href {\doibase 10.1103/PhysRevC.78.028801} {\bibfield
  {journal} {\bibinfo  {journal} {\prc}\ }\textbf {\bibinfo {volume} {78}},\
  \bibinfo {eid} {028801} (\bibinfo {year} {2008})}\BibitemShut {NoStop}%
\bibitem [{\citenamefont {{Burgio}}\ and\ \citenamefont
  {{Schulze}}(2010)}]{Burgio10}%
  \BibitemOpen
  \bibfield  {author} {\bibinfo {author} {\bibfnamefont {G.~F.}\ \bibnamefont
  {{Burgio}}}\ and\ \bibinfo {author} {\bibfnamefont {H.-J.}\ \bibnamefont
  {{Schulze}}},\ }\href {\doibase 10.1051/0004-6361/201014308} {\bibfield
  {journal} {\bibinfo  {journal} {\aap}\ }\textbf {\bibinfo {volume} {518}},\
  \bibinfo {eid} {A17} (\bibinfo {year} {2010})}\BibitemShut {NoStop}%
\bibitem [{\citenamefont {{Lu}}\ \emph {et~al.}(2019)\citenamefont {{Lu}},
  \citenamefont {{Li}}, \citenamefont {{Burgio}}, \citenamefont {{Figura}},\
  and\ \citenamefont {{Schulze}}}]{Lu19}%
  \BibitemOpen
  \bibfield  {author} {\bibinfo {author} {\bibfnamefont {J.-J.}\ \bibnamefont
  {{Lu}}}, \bibinfo {author} {\bibfnamefont {Z.-H.}\ \bibnamefont {{Li}}},
  \bibinfo {author} {\bibfnamefont {G.~F.}\ \bibnamefont {{Burgio}}}, \bibinfo
  {author} {\bibfnamefont {A.}~\bibnamefont {{Figura}}}, \ and\ \bibinfo
  {author} {\bibfnamefont {H.-J.}\ \bibnamefont {{Schulze}}},\ }\href {\doibase
  10.1103/PhysRevC.100.054335} {\bibfield  {journal} {\bibinfo  {journal}
  {\prc}\ }\textbf {\bibinfo {volume} {100}},\ \bibinfo {eid} {054335}
  (\bibinfo {year} {2019})}\BibitemShut {NoStop}%
\bibitem [{\citenamefont {{Shen}}\ \emph {et~al.}(1998)\citenamefont {{Shen}},
  \citenamefont {{Toki}}, \citenamefont {{Oyamatsu}},\ and\ \citenamefont
  {{Sumiyoshi}}}]{1998PThPh.100.1013S}%
  \BibitemOpen
  \bibfield  {author} {\bibinfo {author} {\bibfnamefont {H.}~\bibnamefont
  {{Shen}}}, \bibinfo {author} {\bibfnamefont {H.}~\bibnamefont {{Toki}}},
  \bibinfo {author} {\bibfnamefont {K.}~\bibnamefont {{Oyamatsu}}}, \ and\
  \bibinfo {author} {\bibfnamefont {K.}~\bibnamefont {{Sumiyoshi}}},\ }\href
  {\doibase 10.1143/PTP.100.1013} {\bibfield  {journal} {\bibinfo  {journal}
  {Prog. Theor. Phys.}\ }\textbf {\bibinfo {volume} {100}},\ \bibinfo {pages}
  {1013} (\bibinfo {year} {1998})}\BibitemShut {NoStop}%
\bibitem [{\citenamefont {{Shen}}\ \emph {et~al.}(2011)\citenamefont {{Shen}},
  \citenamefont {{Toki}}, \citenamefont {{Oyamatsu}},\ and\ \citenamefont
  {{Sumiyoshi}}}]{Shen11}%
  \BibitemOpen
  \bibfield  {author} {\bibinfo {author} {\bibfnamefont {H.}~\bibnamefont
  {{Shen}}}, \bibinfo {author} {\bibfnamefont {H.}~\bibnamefont {{Toki}}},
  \bibinfo {author} {\bibfnamefont {K.}~\bibnamefont {{Oyamatsu}}}, \ and\
  \bibinfo {author} {\bibfnamefont {K.}~\bibnamefont {{Sumiyoshi}}},\ }\href
  {\doibase 10.1088/0067-0049/197/2/20} {\bibfield  {journal} {\bibinfo
  {journal} {\apjs}\ }\textbf {\bibinfo {volume} {197}},\ \bibinfo {eid} {20}
  (\bibinfo {year} {2011})}\BibitemShut {NoStop}%
\bibitem [{\citenamefont {Sugahara}\ and\ \citenamefont {Toki}(1994)}]{TM1}%
  \BibitemOpen
  \bibfield  {author} {\bibinfo {author} {\bibfnamefont {Y.}~\bibnamefont
  {Sugahara}}\ and\ \bibinfo {author} {\bibfnamefont {H.}~\bibnamefont
  {Toki}},\ }\href {\doibase https://doi.org/10.1016/0375-9474(94)90923-7}
  {\bibfield  {journal} {\bibinfo  {journal} {\npa}\ }\textbf {\bibinfo
  {volume} {579}},\ \bibinfo {pages} {557} (\bibinfo {year}
  {1994})}\BibitemShut {NoStop}%
\bibitem [{\citenamefont {{Shen}}\ \emph {et~al.}(2020)\citenamefont {{Shen}},
  \citenamefont {{Ji}}, \citenamefont {{Hu}},\ and\ \citenamefont
  {{Sumiyoshi}}}]{Shen20}%
  \BibitemOpen
  \bibfield  {author} {\bibinfo {author} {\bibfnamefont {H.}~\bibnamefont
  {{Shen}}}, \bibinfo {author} {\bibfnamefont {F.}~\bibnamefont {{Ji}}},
  \bibinfo {author} {\bibfnamefont {J.}~\bibnamefont {{Hu}}}, \ and\ \bibinfo
  {author} {\bibfnamefont {K.}~\bibnamefont {{Sumiyoshi}}},\ }\href@noop {}
  {\bibfield  {journal} {\bibinfo  {journal} {\apj}\ }\textbf {\bibinfo
  {volume} {891}},\ \bibinfo {eid} {148} (\bibinfo {year} {2020})}\BibitemShut
  {NoStop}%
\bibitem [{\citenamefont {Bao}\ \emph {et~al.}(2014)\citenamefont {Bao},
  \citenamefont {Hu}, \citenamefont {Zhang},\ and\ \citenamefont
  {Shen}}]{TM1e}%
  \BibitemOpen
  \bibfield  {author} {\bibinfo {author} {\bibfnamefont {S.~S.}\ \bibnamefont
  {Bao}}, \bibinfo {author} {\bibfnamefont {J.~N.}\ \bibnamefont {Hu}},
  \bibinfo {author} {\bibfnamefont {Z.~W.}\ \bibnamefont {Zhang}}, \ and\
  \bibinfo {author} {\bibfnamefont {H.}~\bibnamefont {Shen}},\ }\href {\doibase
  10.1103/PhysRevC.90.045802} {\bibfield  {journal} {\bibinfo  {journal} {Phys.
  Rev. C}\ }\textbf {\bibinfo {volume} {90}},\ \bibinfo {pages} {045802}
  (\bibinfo {year} {2014})}\BibitemShut {NoStop}%
\bibitem [{\citenamefont {Hempel}\ and\ \citenamefont
  {Schaffner-Bielich}(2010)}]{Hempel10}%
  \BibitemOpen
  \bibfield  {author} {\bibinfo {author} {\bibfnamefont {M.}~\bibnamefont
  {Hempel}}\ and\ \bibinfo {author} {\bibfnamefont {J.}~\bibnamefont
  {Schaffner-Bielich}},\ }\href@noop {} {\bibfield  {journal} {\bibinfo
  {journal} {Nucl. Phys. A}\ }\textbf {\bibinfo {volume} {837}},\ \bibinfo
  {pages} {210} (\bibinfo {year} {2010})}\BibitemShut {NoStop}%
\bibitem [{\citenamefont {Typel}\ \emph {et~al.}(2010)\citenamefont {Typel},
  \citenamefont {R\"opke}, \citenamefont {Kl\"ahn}, \citenamefont {Blaschke},\
  and\ \citenamefont {Wolter}}]{DD2}%
  \BibitemOpen
  \bibfield  {author} {\bibinfo {author} {\bibfnamefont {S.}~\bibnamefont
  {Typel}}, \bibinfo {author} {\bibfnamefont {G.}~\bibnamefont {R\"opke}},
  \bibinfo {author} {\bibfnamefont {T.}~\bibnamefont {Kl\"ahn}}, \bibinfo
  {author} {\bibfnamefont {D.}~\bibnamefont {Blaschke}}, \ and\ \bibinfo
  {author} {\bibfnamefont {H.~H.}\ \bibnamefont {Wolter}},\ }\href {\doibase
  10.1103/PhysRevC.81.015803} {\bibfield  {journal} {\bibinfo  {journal} {Phys.
  Rev. C}\ }\textbf {\bibinfo {volume} {81}},\ \bibinfo {pages} {015803}
  (\bibinfo {year} {2010})}\BibitemShut {NoStop}%
\bibitem [{\citenamefont {{Steiner}}\ \emph {et~al.}(2013)\citenamefont
  {{Steiner}}, \citenamefont {{Hempel}},\ and\ \citenamefont
  {{Fischer}}}]{Steiner13}%
  \BibitemOpen
  \bibfield  {author} {\bibinfo {author} {\bibfnamefont {A.~W.}\ \bibnamefont
  {{Steiner}}}, \bibinfo {author} {\bibfnamefont {M.}~\bibnamefont {{Hempel}}},
  \ and\ \bibinfo {author} {\bibfnamefont {T.}~\bibnamefont {{Fischer}}},\
  }\href {\doibase 10.1088/0004-637X/774/1/17} {\bibfield  {journal} {\bibinfo
  {journal} {\apj}\ }\textbf {\bibinfo {volume} {774}},\ \bibinfo {eid} {17}
  (\bibinfo {year} {2013})}\BibitemShut {NoStop}%
\bibitem [{\citenamefont {Tolos}\ \emph
  {et~al.}(2017{\natexlab{a}})\citenamefont {Tolos}, \citenamefont
  {Centelles},\ and\ \citenamefont {Ramos}}]{Tolos17}%
  \BibitemOpen
  \bibfield  {author} {\bibinfo {author} {\bibfnamefont {L.}~\bibnamefont
  {Tolos}}, \bibinfo {author} {\bibfnamefont {M.}~\bibnamefont {Centelles}}, \
  and\ \bibinfo {author} {\bibfnamefont {A.}~\bibnamefont {Ramos}},\
  }\href@noop {} {\bibfield  {journal} {\bibinfo  {journal} {Astrophys. J.}\
  }\textbf {\bibinfo {volume} {834}},\ \bibinfo {pages} {3} (\bibinfo {year}
  {2017}{\natexlab{a}})}\BibitemShut {NoStop}%
\bibitem [{\citenamefont {Tolos}\ \emph
  {et~al.}(2017{\natexlab{b}})\citenamefont {Tolos}, \citenamefont
  {Centelles},\ and\ \citenamefont {Ramos}}]{Tolos17b}%
  \BibitemOpen
  \bibfield  {author} {\bibinfo {author} {\bibfnamefont {L.}~\bibnamefont
  {Tolos}}, \bibinfo {author} {\bibfnamefont {M.}~\bibnamefont {Centelles}}, \
  and\ \bibinfo {author} {\bibfnamefont {A.}~\bibnamefont {Ramos}},\
  }\href@noop {} {\bibfield  {journal} {\bibinfo  {journal} {Publ. Astron. Soc.
  Austral.}\ }\textbf {\bibinfo {volume} {34}},\ \bibinfo {pages} {e065}
  (\bibinfo {year} {2017}{\natexlab{b}})}\BibitemShut {NoStop}%
\bibitem [{\citenamefont {Kanzawa}\ \emph {et~al.}(2007)\citenamefont
  {Kanzawa}, \citenamefont {Oyamatsu}, \citenamefont {Sumiyoshi},\ and\
  \citenamefont {Takano}}]{Kanzawa07}%
  \BibitemOpen
  \bibfield  {author} {\bibinfo {author} {\bibfnamefont {H.}~\bibnamefont
  {Kanzawa}}, \bibinfo {author} {\bibfnamefont {K.}~\bibnamefont {Oyamatsu}},
  \bibinfo {author} {\bibfnamefont {K.}~\bibnamefont {Sumiyoshi}}, \ and\
  \bibinfo {author} {\bibfnamefont {M.}~\bibnamefont {Takano}},\ }\href@noop {}
  {\bibfield  {journal} {\bibinfo  {journal} {\npa}\ }\textbf {\bibinfo
  {volume} {791}},\ \bibinfo {pages} {232} (\bibinfo {year}
  {2007})}\BibitemShut {NoStop}%
\bibitem [{\citenamefont {Typel}\ \emph {et~al.}(2015)\citenamefont {Typel},
  \citenamefont {Oertel},\ and\ \citenamefont {{Kl\"ahn}}}]{Typel_2013}%
  \BibitemOpen
  \bibfield  {author} {\bibinfo {author} {\bibfnamefont {S.}~\bibnamefont
  {Typel}}, \bibinfo {author} {\bibfnamefont {M.}~\bibnamefont {Oertel}}, \
  and\ \bibinfo {author} {\bibfnamefont {T.}~\bibnamefont {{Kl\"ahn}}},\ }\href
  {\doibase 10.1134/S1063779615040061} {\bibfield  {journal} {\bibinfo
  {journal} {Phys. Part. Nucl.}\ }\textbf {\bibinfo {volume} {46}},\ \bibinfo
  {pages} {633} (\bibinfo {year} {2015})}\BibitemShut {NoStop}%
\bibitem [{Com()}]{CompOSE}%
  \BibitemOpen
  \href {\doibase https://compose.obspm.fr} {\
  https://compose.obspm.fr}\BibitemShut {NoStop}%
\bibitem [{BHF()}]{BHFSUP}%
  \BibitemOpen
  \href {\doibase
  https://journals.aps.org/prc/abstract/10.1103/PhysRevC.100.054335$\#$supplemental}
  {\
  https://journals.aps.org/prc/abstract/10.1103/PhysRevC.100.054335$\#$supplemental}\BibitemShut
  {NoStop}%
\bibitem [{She()}]{ShenWWW}%
  \BibitemOpen
  \href {\doibase https://user.numazu-ct.ac.jp/$\sim$sumi/eos} {\
  https://user.numazu-ct.ac.jp/$\sim$sumi/eos}\BibitemShut {NoStop}%
\bibitem [{\citenamefont {{Lattimer}}\ and\ \citenamefont
  {{Swesty}}(1991)}]{Lattimer91}%
  \BibitemOpen
  \bibfield  {author} {\bibinfo {author} {\bibfnamefont {J.~M.}\ \bibnamefont
  {{Lattimer}}}\ and\ \bibinfo {author} {\bibfnamefont {D.~F.}\ \bibnamefont
  {{Swesty}}},\ }\href {\doibase 10.1016/0375-9474(91)90452-C} {\bibfield
  {journal} {\bibinfo  {journal} {\nphysa}\ }\textbf {\bibinfo {volume}
  {535}},\ \bibinfo {pages} {331} (\bibinfo {year} {1991})}\BibitemShut
  {NoStop}%
\bibitem [{LSW()}]{LSWWW}%
  \BibitemOpen
  \href {\doibase http://www.astro.sunysb.edu/dswesty/lseos.html} {\
  http://www.astro.sunysb.edu/dswesty/lseos.html}\BibitemShut {NoStop}%
\bibitem [{\citenamefont {{Kaplan}}\ \emph {et~al.}(2014)\citenamefont
  {{Kaplan}}, \citenamefont {{Ott}}, \citenamefont {{O'Connor}}, \citenamefont
  {{Kiuchi}}, \citenamefont {{Roberts}},\ and\ \citenamefont
  {{Duez}}}]{Kaplan14}%
  \BibitemOpen
  \bibfield  {author} {\bibinfo {author} {\bibfnamefont {J.~D.}\ \bibnamefont
  {{Kaplan}}}, \bibinfo {author} {\bibfnamefont {C.~D.}\ \bibnamefont {{Ott}}},
  \bibinfo {author} {\bibfnamefont {E.~P.}\ \bibnamefont {{O'Connor}}},
  \bibinfo {author} {\bibfnamefont {K.}~\bibnamefont {{Kiuchi}}}, \bibinfo
  {author} {\bibfnamefont {L.}~\bibnamefont {{Roberts}}}, \ and\ \bibinfo
  {author} {\bibfnamefont {M.}~\bibnamefont {{Duez}}},\ }\href {\doibase
  10.1088/0004-637X/790/1/19} {\bibfield  {journal} {\bibinfo  {journal}
  {\apj}\ }\textbf {\bibinfo {volume} {790}},\ \bibinfo {eid} {19} (\bibinfo
  {year} {2014})}\BibitemShut {NoStop}%
\bibitem [{\citenamefont {{da Silva Schneider}}\ \emph
  {et~al.}(2020)\citenamefont {{da Silva Schneider}}, \citenamefont
  {{O'Connor}}, \citenamefont {{Granqvist}}, \citenamefont {{Betranhandy}},\
  and\ \citenamefont {{Couch}}}]{Schneider20}%
  \BibitemOpen
  \bibfield  {author} {\bibinfo {author} {\bibfnamefont {A.}~\bibnamefont {{da
  Silva Schneider}}}, \bibinfo {author} {\bibfnamefont {E.}~\bibnamefont
  {{O'Connor}}}, \bibinfo {author} {\bibfnamefont {E.}~\bibnamefont
  {{Granqvist}}}, \bibinfo {author} {\bibfnamefont {A.}~\bibnamefont
  {{Betranhandy}}}, \ and\ \bibinfo {author} {\bibfnamefont {S.~M.}\
  \bibnamefont {{Couch}}},\ }\href {\doibase 10.3847/1538-4357/ab8308}
  {\bibfield  {journal} {\bibinfo  {journal} {\apj}\ }\textbf {\bibinfo
  {volume} {894}},\ \bibinfo {eid} {4} (\bibinfo {year} {2020})}\BibitemShut
  {NoStop}%
\bibitem [{\citenamefont {{Raduta}}\ \emph {et~al.}(2020)\citenamefont
  {{Raduta}}, \citenamefont {{Oertel}},\ and\ \citenamefont
  {{Sedrakian}}}]{Raduta20}%
  \BibitemOpen
  \bibfield  {author} {\bibinfo {author} {\bibfnamefont {A.~R.}\ \bibnamefont
  {{Raduta}}}, \bibinfo {author} {\bibfnamefont {M.}~\bibnamefont {{Oertel}}},
  \ and\ \bibinfo {author} {\bibfnamefont {A.}~\bibnamefont {{Sedrakian}}},\
  }\href {\doibase 10.1093/mnras/staa2491} {\bibfield  {journal} {\bibinfo
  {journal} {\mnras}\ }\textbf {\bibinfo {volume} {499}},\ \bibinfo {pages}
  {914} (\bibinfo {year} {2020})}\BibitemShut {NoStop}%
\bibitem [{\citenamefont {{Baumgarte}}\ \emph {et~al.}(1995)\citenamefont
  {{Baumgarte}}, \citenamefont {{Shapiro}},\ and\ \citenamefont
  {{Teukolsky}}}]{Baumgarte95}%
  \BibitemOpen
  \bibfield  {author} {\bibinfo {author} {\bibfnamefont {T.~W.}\ \bibnamefont
  {{Baumgarte}}}, \bibinfo {author} {\bibfnamefont {S.~L.}\ \bibnamefont
  {{Shapiro}}}, \ and\ \bibinfo {author} {\bibfnamefont {S.~A.}\ \bibnamefont
  {{Teukolsky}}},\ }\href {\doibase 10.1086/175563} {\bibfield  {journal}
  {\bibinfo  {journal} {\apj}\ }\textbf {\bibinfo {volume} {443}},\ \bibinfo
  {pages} {717} (\bibinfo {year} {1995})}\BibitemShut {NoStop}%
\bibitem [{\citenamefont {{Prakash}}\ \emph {et~al.}(1997)\citenamefont
  {{Prakash}}, \citenamefont {{Bombaci}}, \citenamefont {{Prakash}},
  \citenamefont {{Ellis}}, \citenamefont {{Lattimer}},\ and\ \citenamefont
  {{Knorren}}}]{Prakash97}%
  \BibitemOpen
  \bibfield  {author} {\bibinfo {author} {\bibfnamefont {M.}~\bibnamefont
  {{Prakash}}}, \bibinfo {author} {\bibfnamefont {I.}~\bibnamefont
  {{Bombaci}}}, \bibinfo {author} {\bibfnamefont {M.}~\bibnamefont
  {{Prakash}}}, \bibinfo {author} {\bibfnamefont {P.~J.}\ \bibnamefont
  {{Ellis}}}, \bibinfo {author} {\bibfnamefont {J.~M.}\ \bibnamefont
  {{Lattimer}}}, \ and\ \bibinfo {author} {\bibfnamefont {R.}~\bibnamefont
  {{Knorren}}},\ }\href {\doibase 10.1016/S0370-1573(96)00023-3} {\bibfield
  {journal} {\bibinfo  {journal} {\physrep}\ }\textbf {\bibinfo {volume}
  {280}},\ \bibinfo {pages} {1} (\bibinfo {year} {1997})}\BibitemShut {NoStop}%
\bibitem [{\citenamefont {{Nicotra}}\ \emph {et~al.}(2006)\citenamefont
  {{Nicotra}}, \citenamefont {{Baldo}}, \citenamefont {{Burgio}},\ and\
  \citenamefont {{Schulze}}}]{Nicotra06}%
  \BibitemOpen
  \bibfield  {author} {\bibinfo {author} {\bibfnamefont {O.~E.}\ \bibnamefont
  {{Nicotra}}}, \bibinfo {author} {\bibfnamefont {M.}~\bibnamefont {{Baldo}}},
  \bibinfo {author} {\bibfnamefont {G.~F.}\ \bibnamefont {{Burgio}}}, \ and\
  \bibinfo {author} {\bibfnamefont {H.-J.}\ \bibnamefont {{Schulze}}},\ }\href
  {\doibase 10.1051/0004-6361:20053670} {\bibfield  {journal} {\bibinfo
  {journal} {\aap}\ }\textbf {\bibinfo {volume} {451}},\ \bibinfo {pages} {213}
  (\bibinfo {year} {2006})}\BibitemShut {NoStop}%
\bibitem [{\citenamefont {{Li}}\ \emph {et~al.}(2010)\citenamefont {{Li}},
  \citenamefont {{Zhou}}, \citenamefont {{Burgio}},\ and\ \citenamefont
  {{Schulze}}}]{Li10}%
  \BibitemOpen
  \bibfield  {author} {\bibinfo {author} {\bibfnamefont {A.}~\bibnamefont
  {{Li}}}, \bibinfo {author} {\bibfnamefont {X.~R.}\ \bibnamefont {{Zhou}}},
  \bibinfo {author} {\bibfnamefont {G.~F.}\ \bibnamefont {{Burgio}}}, \ and\
  \bibinfo {author} {\bibfnamefont {H.-J.}\ \bibnamefont {{Schulze}}},\ }\href
  {\doibase 10.1103/PhysRevC.81.025806} {\bibfield  {journal} {\bibinfo
  {journal} {\prc}\ }\textbf {\bibinfo {volume} {81}},\ \bibinfo {eid} {025806}
  (\bibinfo {year} {2010})}\BibitemShut {NoStop}%
\bibitem [{\citenamefont {{Figura}}\ \emph {et~al.}(2020)\citenamefont
  {{Figura}}, \citenamefont {{Lu}}, \citenamefont {{Burgio}}, \citenamefont
  {{Li}},\ and\ \citenamefont {{Schulze}}}]{Figura20}%
  \BibitemOpen
  \bibfield  {author} {\bibinfo {author} {\bibfnamefont {A.}~\bibnamefont
  {{Figura}}}, \bibinfo {author} {\bibfnamefont {J.-J.}\ \bibnamefont {{Lu}}},
  \bibinfo {author} {\bibfnamefont {G.~F.}\ \bibnamefont {{Burgio}}}, \bibinfo
  {author} {\bibfnamefont {Z.-H.}\ \bibnamefont {{Li}}}, \ and\ \bibinfo
  {author} {\bibfnamefont {H.-J.}\ \bibnamefont {{Schulze}}},\ }\href {\doibase
  10.1103/PhysRevD.102.043006} {\bibfield  {journal} {\bibinfo  {journal}
  {\prd}\ }\textbf {\bibinfo {volume} {102}},\ \bibinfo {eid} {043006}
  (\bibinfo {year} {2020})}\BibitemShut {NoStop}%
\bibitem [{\citenamefont {{Figura}}\ \emph {et~al.}(2021)\citenamefont
  {{Figura}}, \citenamefont {{Li}}, \citenamefont {{Lu}}, \citenamefont
  {{Burgio}}, \citenamefont {{Li}},\ and\ \citenamefont
  {{Schulze}}}]{Figura21}%
  \BibitemOpen
  \bibfield  {author} {\bibinfo {author} {\bibfnamefont {A.}~\bibnamefont
  {{Figura}}}, \bibinfo {author} {\bibfnamefont {F.}~\bibnamefont {{Li}}},
  \bibinfo {author} {\bibfnamefont {J.-J.}\ \bibnamefont {{Lu}}}, \bibinfo
  {author} {\bibfnamefont {G.~F.}\ \bibnamefont {{Burgio}}}, \bibinfo {author}
  {\bibfnamefont {Z.-H.}\ \bibnamefont {{Li}}}, \ and\ \bibinfo {author}
  {\bibfnamefont {H.-J.}\ \bibnamefont {{Schulze}}},\ }\href {\doibase
  10.1103/PhysRevD.103.083012} {\bibfield  {journal} {\bibinfo  {journal}
  {\prd}\ }\textbf {\bibinfo {volume} {103}},\ \bibinfo {eid} {083012}
  (\bibinfo {year} {2021})}\BibitemShut {NoStop}%
\bibitem [{\citenamefont {{Shapiro}}\ and\ \citenamefont
  {{Teukolsky}}(2008)}]{Shapiro08}%
  \BibitemOpen
  \bibfield  {author} {\bibinfo {author} {\bibfnamefont {S.~L.}\ \bibnamefont
  {{Shapiro}}}\ and\ \bibinfo {author} {\bibfnamefont {S.~A.}\ \bibnamefont
  {{Teukolsky}}},\ }\href@noop {} {\emph {\bibinfo {title} {Black Holes, White
  Dwarfs and Neutron Stars: The Physics of Compact Objects}}}\ (\bibinfo
  {publisher} {Wiley},\ \bibinfo {year} {2008})\BibitemShut {NoStop}%
\bibitem [{\citenamefont {{Sun}}\ \emph {et~al.}(2021)\citenamefont {{Sun}},
  \citenamefont {{Zheng}}, \citenamefont {{Chen}}, \citenamefont {{Burgio}},\
  and\ \citenamefont {{Schulze}}}]{Sun21}%
  \BibitemOpen
  \bibfield  {author} {\bibinfo {author} {\bibfnamefont {T.-T.}\ \bibnamefont
  {{Sun}}}, \bibinfo {author} {\bibfnamefont {Z.-Y.}\ \bibnamefont {{Zheng}}},
  \bibinfo {author} {\bibfnamefont {H.}~\bibnamefont {{Chen}}}, \bibinfo
  {author} {\bibfnamefont {G.~F.}\ \bibnamefont {{Burgio}}}, \ and\ \bibinfo
  {author} {\bibfnamefont {H.-J.}\ \bibnamefont {{Schulze}}},\ }\href@noop {}
  {\bibfield  {journal} {\bibinfo  {journal} {\prd}\ }\textbf {\bibinfo
  {volume} {103}},\ \bibinfo {eid} {103003} (\bibinfo {year}
  {2021})}\BibitemShut {NoStop}%
\bibitem [{\citenamefont {{H. T. Cromartie {\it et al.}}}(2020)}]{Cromartie20}%
  \BibitemOpen
  \bibfield  {author} {\bibinfo {author} {\bibnamefont {{H. T. Cromartie {\it
  et al.}}}},\ }\href@noop {} {\bibfield  {journal} {\bibinfo  {journal}
  {Nature Astronomy}\ }\textbf {\bibinfo {volume} {4}},\ \bibinfo {pages} {72}
  (\bibinfo {year} {2020})}\BibitemShut {NoStop}%
\bibitem [{\citenamefont {{Pang}}\ \emph {et~al.}(2021)\citenamefont {{Pang}},
  \citenamefont {{Tews}}, \citenamefont {{Coughlin}}, \citenamefont {{Bulla}},
  \citenamefont {{Van Den Broeck}},\ and\ \citenamefont {{Dietrich}}}]{Pang21}%
  \BibitemOpen
  \bibfield  {author} {\bibinfo {author} {\bibfnamefont {P.~T.~H.}\
  \bibnamefont {{Pang}}}, \bibinfo {author} {\bibfnamefont {I.}~\bibnamefont
  {{Tews}}}, \bibinfo {author} {\bibfnamefont {M.~W.}\ \bibnamefont
  {{Coughlin}}}, \bibinfo {author} {\bibfnamefont {M.}~\bibnamefont {{Bulla}}},
  \bibinfo {author} {\bibfnamefont {C.}~\bibnamefont {{Van Den Broeck}}}, \
  and\ \bibinfo {author} {\bibfnamefont {T.}~\bibnamefont {{Dietrich}}},\
  }\href@noop {} {\ ,\ \bibinfo {eid} {arXiv:2105.08688} (\bibinfo {year}
  {2021})}\BibitemShut {NoStop}%
\bibitem [{\citenamefont {{Raaijmakers}}\ \emph {et~al.}(2021)\citenamefont
  {{Raaijmakers}}, \citenamefont {{Greif}}, \citenamefont {{Hebeler}},
  \citenamefont {{Hinderer}}, \citenamefont {{Nissanke}}, \citenamefont
  {{Schwenk}}, \citenamefont {{Riley}}, \citenamefont {{Watts}}, \citenamefont
  {{Lattimer}},\ and\ \citenamefont {{Ho}}}]{Raaijmakers21}%
  \BibitemOpen
  \bibfield  {author} {\bibinfo {author} {\bibfnamefont {G.}~\bibnamefont
  {{Raaijmakers}}}, \bibinfo {author} {\bibfnamefont {S.~K.}\ \bibnamefont
  {{Greif}}}, \bibinfo {author} {\bibfnamefont {K.}~\bibnamefont {{Hebeler}}},
  \bibinfo {author} {\bibfnamefont {T.}~\bibnamefont {{Hinderer}}}, \bibinfo
  {author} {\bibfnamefont {S.}~\bibnamefont {{Nissanke}}}, \bibinfo {author}
  {\bibfnamefont {A.}~\bibnamefont {{Schwenk}}}, \bibinfo {author}
  {\bibfnamefont {T.~E.}\ \bibnamefont {{Riley}}}, \bibinfo {author}
  {\bibfnamefont {A.~L.}\ \bibnamefont {{Watts}}}, \bibinfo {author}
  {\bibfnamefont {J.~M.}\ \bibnamefont {{Lattimer}}}, \ and\ \bibinfo {author}
  {\bibfnamefont {W.~C.~G.}\ \bibnamefont {{Ho}}},\ }\href@noop {} {\bibfield
  {journal} {\bibinfo  {journal} {\apjl}\ }\textbf {\bibinfo {volume} {918}},\
  \bibinfo {eid} {L29} (\bibinfo {year} {2021})}\BibitemShut {NoStop}%
\bibitem [{\citenamefont {Lu}\ \emph {et~al.}(2019)\citenamefont {Lu},
  \citenamefont {Li}, \citenamefont {Burgio}, \citenamefont {Figura},\ and\
  \citenamefont {Schulze}}]{BHFWWW}%
  \BibitemOpen
  \bibfield  {author} {\bibinfo {author} {\bibfnamefont {J.-J.}\ \bibnamefont
  {Lu}}, \bibinfo {author} {\bibfnamefont {Z.-H.}\ \bibnamefont {Li}}, \bibinfo
  {author} {\bibfnamefont {G.~F.}\ \bibnamefont {Burgio}}, \bibinfo {author}
  {\bibfnamefont {A.}~\bibnamefont {Figura}}, \ and\ \bibinfo {author}
  {\bibfnamefont {H.-J.}\ \bibnamefont {Schulze}},\ }\href {\doibase
  10.1103/PhysRevC.100.054335} {\bibfield  {journal} {\bibinfo  {journal}
  {Phys. Rev. C}\ }\textbf {\bibinfo {volume} {100}},\ \bibinfo {pages}
  {054335} (\bibinfo {year} {2019})}\BibitemShut {NoStop}%
\bibitem [{\citenamefont {{Margueron}}\ \emph {et~al.}(2018)\citenamefont
  {{Margueron}}, \citenamefont {{Hoffmann Casali}},\ and\ \citenamefont
  {{Gulminelli}}}]{Margueron18a}%
  \BibitemOpen
  \bibfield  {author} {\bibinfo {author} {\bibfnamefont {J.}~\bibnamefont
  {{Margueron}}}, \bibinfo {author} {\bibfnamefont {R.}~\bibnamefont {{Hoffmann
  Casali}}}, \ and\ \bibinfo {author} {\bibfnamefont {F.}~\bibnamefont
  {{Gulminelli}}},\ }\href@noop {} {\bibfield  {journal} {\bibinfo  {journal}
  {\prc}\ }\textbf {\bibinfo {volume} {97}},\ \bibinfo {eid} {025805} (\bibinfo
  {year} {2018})}\BibitemShut {NoStop}%
\bibitem [{\citenamefont {{Shlomo}}\ \emph {et~al.}(2006)\citenamefont
  {{Shlomo}}, \citenamefont {{Kolomietz}},\ and\ \citenamefont
  {{Col{\`o}}}}]{Shlomo06}%
  \BibitemOpen
  \bibfield  {author} {\bibinfo {author} {\bibfnamefont {S.}~\bibnamefont
  {{Shlomo}}}, \bibinfo {author} {\bibfnamefont {V.~M.}\ \bibnamefont
  {{Kolomietz}}}, \ and\ \bibinfo {author} {\bibfnamefont {G.}~\bibnamefont
  {{Col{\`o}}}},\ }\href@noop {} {\bibfield  {journal} {\bibinfo  {journal}
  {\epja}\ }\textbf {\bibinfo {volume} {30}},\ \bibinfo {pages} {23} (\bibinfo
  {year} {2006})}\BibitemShut {NoStop}%
\bibitem [{\citenamefont {{Piekarewicz}}(2010)}]{Piekarewicz10}%
  \BibitemOpen
  \bibfield  {author} {\bibinfo {author} {\bibfnamefont {J.}~\bibnamefont
  {{Piekarewicz}}},\ }\href@noop {} {\bibfield  {journal} {\bibinfo  {journal}
  {J. Phys. G}\ }\textbf {\bibinfo {volume} {37}},\ \bibinfo {eid} {064038}
  (\bibinfo {year} {2010})}\BibitemShut {NoStop}%
\bibitem [{\citenamefont {{Li}}\ and\ \citenamefont {{Han}}(2013)}]{Li13}%
  \BibitemOpen
  \bibfield  {author} {\bibinfo {author} {\bibfnamefont {B.-A.}\ \bibnamefont
  {{Li}}}\ and\ \bibinfo {author} {\bibfnamefont {X.}~\bibnamefont {{Han}}},\
  }\href@noop {} {\bibfield  {journal} {\bibinfo  {journal} {\plb}\ }\textbf
  {\bibinfo {volume} {727}},\ \bibinfo {pages} {276} (\bibinfo {year}
  {2013})}\BibitemShut {NoStop}%
\bibitem [{\citenamefont {{B. Margalit and B. D. Metzger}}(2017)}]{Margalit17}%
  \BibitemOpen
  \bibfield  {author} {\bibinfo {author} {\bibnamefont {{B. Margalit and B. D.
  Metzger}}},\ }\href@noop {} {\bibfield  {journal} {\bibinfo  {journal}
  {Astrophys. J}\ }\textbf {\bibinfo {volume} {850}},\ \bibinfo {pages} {L19}
  (\bibinfo {year} {2017})}\BibitemShut {NoStop}%
\bibitem [{\citenamefont {{Ruiz}}\ \emph {et~al.}(2018)\citenamefont {{Ruiz}},
  \citenamefont {{Shapiro}},\ and\ \citenamefont {{Tsokaros}}}]{Ruiz18}%
  \BibitemOpen
  \bibfield  {author} {\bibinfo {author} {\bibfnamefont {M.}~\bibnamefont
  {{Ruiz}}}, \bibinfo {author} {\bibfnamefont {S.~L.}\ \bibnamefont
  {{Shapiro}}}, \ and\ \bibinfo {author} {\bibfnamefont {A.}~\bibnamefont
  {{Tsokaros}}},\ }\href {\doibase 10.1103/PhysRevD.97.021501} {\bibfield
  {journal} {\bibinfo  {journal} {\prd}\ }\textbf {\bibinfo {volume} {97}},\
  \bibinfo {eid} {021501} (\bibinfo {year} {2018})}\BibitemShut {NoStop}%
\bibitem [{\citenamefont {Most}\ \emph {et~al.}(2018)\citenamefont {Most},
  \citenamefont {Weih}, \citenamefont {Rezzolla},\ and\ \citenamefont
  {Schaffner-Bielich}}]{Most18}%
  \BibitemOpen
  \bibfield  {author} {\bibinfo {author} {\bibfnamefont {E.~R.}\ \bibnamefont
  {Most}}, \bibinfo {author} {\bibfnamefont {L.~R.}\ \bibnamefont {Weih}},
  \bibinfo {author} {\bibfnamefont {L.}~\bibnamefont {Rezzolla}}, \ and\
  \bibinfo {author} {\bibfnamefont {J.}~\bibnamefont {Schaffner-Bielich}},\
  }\href@noop {} {\bibfield  {journal} {\bibinfo  {journal} {Phys. Rev. Lett.}\
  }\textbf {\bibinfo {volume} {120}},\ \bibinfo {pages} {261103} (\bibinfo
  {year} {2018})}\BibitemShut {NoStop}%
\bibitem [{\citenamefont {{L. Rezzolla, E. R. Most, and L. R.
  Weih}}(2018)}]{Rezzolla18}%
  \BibitemOpen
  \bibfield  {author} {\bibinfo {author} {\bibnamefont {{L. Rezzolla, E. R.
  Most, and L. R. Weih}}},\ }\href@noop {} {\bibfield  {journal} {\bibinfo
  {journal} {Astrophys. J}\ }\textbf {\bibinfo {volume} {852}},\ \bibinfo
  {pages} {L25} (\bibinfo {year} {2018})}\BibitemShut {NoStop}%
\bibitem [{\citenamefont {{Shibata}}\ \emph {et~al.}(2019)\citenamefont
  {{Shibata}}, \citenamefont {{Zhou}}, \citenamefont {{Kiuchi}},\ and\
  \citenamefont {{Fujibayashi}}}]{Shibata19}%
  \BibitemOpen
  \bibfield  {author} {\bibinfo {author} {\bibfnamefont {M.}~\bibnamefont
  {{Shibata}}}, \bibinfo {author} {\bibfnamefont {E.}~\bibnamefont {{Zhou}}},
  \bibinfo {author} {\bibfnamefont {K.}~\bibnamefont {{Kiuchi}}}, \ and\
  \bibinfo {author} {\bibfnamefont {S.}~\bibnamefont {{Fujibayashi}}},\
  }\href@noop {} {\bibfield  {journal} {\bibinfo  {journal} {\prd}\ }\textbf
  {\bibinfo {volume} {100}},\ \bibinfo {eid} {023015} (\bibinfo {year}
  {2019})}\BibitemShut {NoStop}%
\bibitem [{\citenamefont {{Most}}\ \emph {et~al.}(2020)\citenamefont {{Most}},
  \citenamefont {{Papenfort}}, \citenamefont {{Weih}},\ and\ \citenamefont
  {{Rezzolla}}}]{Most20}%
  \BibitemOpen
  \bibfield  {author} {\bibinfo {author} {\bibfnamefont {E.~R.}\ \bibnamefont
  {{Most}}}, \bibinfo {author} {\bibfnamefont {L.~J.}\ \bibnamefont
  {{Papenfort}}}, \bibinfo {author} {\bibfnamefont {L.~R.}\ \bibnamefont
  {{Weih}}}, \ and\ \bibinfo {author} {\bibfnamefont {L.}~\bibnamefont
  {{Rezzolla}}},\ }\href {\doibase 10.1093/mnrasl/slaa168} {\bibfield
  {journal} {\bibinfo  {journal} {\mnras}\ }\textbf {\bibinfo {volume} {499}},\
  \bibinfo {pages} {L82} (\bibinfo {year} {2020})}\BibitemShut {NoStop}%
\bibitem [{\citenamefont {{M. Shibata, S. Fujibayashi, K. Hotokezakam, K.
  Kiuchi, K. Kyutoku, Y. Sekiguchi, and M. Tanaka}}(2017)}]{Shibata17}%
  \BibitemOpen
  \bibfield  {author} {\bibinfo {author} {\bibnamefont {{M. Shibata, S.
  Fujibayashi, K. Hotokezakam, K. Kiuchi, K. Kyutoku, Y. Sekiguchi, and M.
  Tanaka}}},\ }\href@noop {} {\bibfield  {journal} {\bibinfo  {journal} {Phys.
  Rev. D}\ }\textbf {\bibinfo {volume} {96}},\ \bibinfo {pages} {123012}
  (\bibinfo {year} {2017})}\BibitemShut {NoStop}%
\bibitem [{\citenamefont {Khadkikar}\ \emph {et~al.}(2021)\citenamefont
  {Khadkikar}, \citenamefont {Raduta}, \citenamefont {Oertel},\ and\
  \citenamefont {Sedrakian}}]{Khadkikar_PRC_2021}%
  \BibitemOpen
  \bibfield  {author} {\bibinfo {author} {\bibfnamefont {S.}~\bibnamefont
  {Khadkikar}}, \bibinfo {author} {\bibfnamefont {A.~R.}\ \bibnamefont
  {Raduta}}, \bibinfo {author} {\bibfnamefont {M.}~\bibnamefont {Oertel}}, \
  and\ \bibinfo {author} {\bibfnamefont {A.}~\bibnamefont {Sedrakian}},\ }\href
  {\doibase 10.1103/PhysRevC.103.055811} {\bibfield  {journal} {\bibinfo
  {journal} {Phys. Rev. C}\ }\textbf {\bibinfo {volume} {103}},\ \bibinfo
  {pages} {055811} (\bibinfo {year} {2021})}\BibitemShut {NoStop}%
\bibitem [{\citenamefont {{Raduta}}\ \emph {et~al.}(2021)\citenamefont
  {{Raduta}}, \citenamefont {{Nacu}},\ and\ \citenamefont
  {{Oertel}}}]{Raduta_2021}%
  \BibitemOpen
  \bibfield  {author} {\bibinfo {author} {\bibfnamefont {A.~R.}\ \bibnamefont
  {{Raduta}}}, \bibinfo {author} {\bibfnamefont {F.}~\bibnamefont {{Nacu}}}, \
  and\ \bibinfo {author} {\bibfnamefont {M.}~\bibnamefont {{Oertel}}},\
  }\href@noop {} {\bibfield  {journal} {\bibinfo  {journal} {arXiv e-prints}\
  ,\ \bibinfo {eid} {arXiv:2109.00251}} (\bibinfo {year} {2021})}\BibitemShut
  {NoStop}%
\bibitem [{\citenamefont {Constantinou}\ \emph {et~al.}(2014)\citenamefont
  {Constantinou}, \citenamefont {Muccioli}, \citenamefont {Prakash},\ and\
  \citenamefont {Lattimer}}]{Constantinou_PRC_2014}%
  \BibitemOpen
  \bibfield  {author} {\bibinfo {author} {\bibfnamefont {C.}~\bibnamefont
  {Constantinou}}, \bibinfo {author} {\bibfnamefont {B.}~\bibnamefont
  {Muccioli}}, \bibinfo {author} {\bibfnamefont {M.}~\bibnamefont {Prakash}}, \
  and\ \bibinfo {author} {\bibfnamefont {J.~M.}\ \bibnamefont {Lattimer}},\
  }\href {\doibase 10.1103/PhysRevC.89.065802} {\bibfield  {journal} {\bibinfo
  {journal} {Phys. Rev. C}\ }\textbf {\bibinfo {volume} {89}},\ \bibinfo
  {pages} {065802} (\bibinfo {year} {2014})}\BibitemShut {NoStop}%
\bibitem [{\citenamefont {{Constantinou}}\ \emph {et~al.}(2015)\citenamefont
  {{Constantinou}}, \citenamefont {{Muccioli}}, \citenamefont {{Prakash}},\
  and\ \citenamefont {{Lattimer}}}]{Constantinou_PRC_2015}%
  \BibitemOpen
  \bibfield  {author} {\bibinfo {author} {\bibfnamefont {C.}~\bibnamefont
  {{Constantinou}}}, \bibinfo {author} {\bibfnamefont {B.}~\bibnamefont
  {{Muccioli}}}, \bibinfo {author} {\bibfnamefont {M.}~\bibnamefont
  {{Prakash}}}, \ and\ \bibinfo {author} {\bibfnamefont {J.~M.}\ \bibnamefont
  {{Lattimer}}},\ }\href {\doibase 10.1103/PhysRevC.92.025801} {\bibfield
  {journal} {\bibinfo  {journal} {\prc}\ }\textbf {\bibinfo {volume} {92}},\
  \bibinfo {eid} {025801} (\bibinfo {year} {2015})}\BibitemShut {NoStop}%
\bibitem [{\citenamefont {{Baldo}}\ \emph {et~al.}(2014)\citenamefont
  {{Baldo}}, \citenamefont {{Burgio}}, \citenamefont {{Schulze}},\ and\
  \citenamefont {{Taranto}}}]{Baldo14}%
  \BibitemOpen
  \bibfield  {author} {\bibinfo {author} {\bibfnamefont {M.}~\bibnamefont
  {{Baldo}}}, \bibinfo {author} {\bibfnamefont {G.~F.}\ \bibnamefont
  {{Burgio}}}, \bibinfo {author} {\bibfnamefont {H.-J.}\ \bibnamefont
  {{Schulze}}}, \ and\ \bibinfo {author} {\bibfnamefont {G.}~\bibnamefont
  {{Taranto}}},\ }\href@noop {} {\bibfield  {journal} {\bibinfo  {journal}
  {\prc}\ }\textbf {\bibinfo {volume} {89}},\ \bibinfo {eid} {048801} (\bibinfo
  {year} {2014})}\BibitemShut {NoStop}%
\bibitem [{\citenamefont {{Raithel}}\ \emph {et~al.}(2019)\citenamefont
  {{Raithel}}, \citenamefont {{{\"O}zel}},\ and\ \citenamefont
  {{Psaltis}}}]{Raithel19}%
  \BibitemOpen
  \bibfield  {author} {\bibinfo {author} {\bibfnamefont {C.~A.}\ \bibnamefont
  {{Raithel}}}, \bibinfo {author} {\bibfnamefont {F.}~\bibnamefont
  {{{\"O}zel}}}, \ and\ \bibinfo {author} {\bibfnamefont {D.}~\bibnamefont
  {{Psaltis}}},\ }\href {\doibase 10.3847/1538-4357/ab08ea} {\bibfield
  {journal} {\bibinfo  {journal} {\apj}\ }\textbf {\bibinfo {volume} {875}},\
  \bibinfo {eid} {12} (\bibinfo {year} {2019})}\BibitemShut {NoStop}%
\bibitem [{\citenamefont {{Raithel}}\ \emph {et~al.}(2021)\citenamefont
  {{Raithel}}, \citenamefont {{{\"O}zel}},\ and\ \citenamefont
  {{Psaltis}}}]{Raithel19_Err}%
  \BibitemOpen
  \bibfield  {author} {\bibinfo {author} {\bibfnamefont {C.~A.}\ \bibnamefont
  {{Raithel}}}, \bibinfo {author} {\bibfnamefont {F.}~\bibnamefont
  {{{\"O}zel}}}, \ and\ \bibinfo {author} {\bibfnamefont {D.}~\bibnamefont
  {{Psaltis}}},\ }\href {\doibase 10.3847/1538-4357/ac0630} {\bibfield
  {journal} {\bibinfo  {journal} {\apj}\ }\textbf {\bibinfo {volume} {915}},\
  \bibinfo {eid} {73} (\bibinfo {year} {2021})}\BibitemShut {NoStop}%
\bibitem [{\citenamefont {Raithel}\ \emph {et~al.}(2021)\citenamefont
  {Raithel}, \citenamefont {Paschalidis},\ and\ \citenamefont
  {{\"O}zel}}]{Raithel21}%
  \BibitemOpen
  \bibfield  {author} {\bibinfo {author} {\bibfnamefont {C.~A.}\ \bibnamefont
  {Raithel}}, \bibinfo {author} {\bibfnamefont {V.}~\bibnamefont
  {Paschalidis}}, \ and\ \bibinfo {author} {\bibfnamefont {F.}~\bibnamefont
  {{\"O}zel}},\ }\href {http://dx.doi.org/10.1103/PhysRevD.104.063016}
  {\bibfield  {journal} {\bibinfo  {journal} {\prd}\ }\textbf {\bibinfo
  {volume} {104}} (\bibinfo {year} {2021})}\BibitemShut {NoStop}%
\bibitem [{\citenamefont {Raduta}\ and\ \citenamefont
  {Gulminelli}(2019)}]{Raduta_NPA_2019}%
  \BibitemOpen
  \bibfield  {author} {\bibinfo {author} {\bibfnamefont {A.}~\bibnamefont
  {Raduta}}\ and\ \bibinfo {author} {\bibfnamefont {F.}~\bibnamefont
  {Gulminelli}},\ }\href@noop {} {\bibfield  {journal} {\bibinfo  {journal}
  {Nucl. Phys. A}\ }\textbf {\bibinfo {volume} {983}},\ \bibinfo {pages} {252}
  (\bibinfo {year} {2019})}\BibitemShut {NoStop}%
\bibitem [{\citenamefont {Chabanat}\ \emph {et~al.}(1998)\citenamefont
  {Chabanat}, \citenamefont {Bonche}, \citenamefont {Haensel}, \citenamefont
  {Meyer},\ and\ \citenamefont {Schaeffer}}]{SLy4}%
  \BibitemOpen
  \bibfield  {author} {\bibinfo {author} {\bibfnamefont {E.}~\bibnamefont
  {Chabanat}}, \bibinfo {author} {\bibfnamefont {P.}~\bibnamefont {Bonche}},
  \bibinfo {author} {\bibfnamefont {P.}~\bibnamefont {Haensel}}, \bibinfo
  {author} {\bibfnamefont {J.}~\bibnamefont {Meyer}}, \ and\ \bibinfo {author}
  {\bibfnamefont {R.}~\bibnamefont {Schaeffer}},\ }\href {\doibase
  https://doi.org/10.1016/S0375-9474(98)00180-8} {\bibfield  {journal}
  {\bibinfo  {journal} {\npa}\ }\textbf {\bibinfo {volume} {635}},\ \bibinfo
  {pages} {231} (\bibinfo {year} {1998})}\BibitemShut {NoStop}%
\bibitem [{\citenamefont {Schneider}\ \emph {et~al.}(2017)\citenamefont
  {Schneider}, \citenamefont {Roberts},\ and\ \citenamefont
  {Ott}}]{SRO_PRC_2017}%
  \BibitemOpen
  \bibfield  {author} {\bibinfo {author} {\bibfnamefont {A.~S.}\ \bibnamefont
  {Schneider}}, \bibinfo {author} {\bibfnamefont {L.~F.}\ \bibnamefont
  {Roberts}}, \ and\ \bibinfo {author} {\bibfnamefont {C.~D.}\ \bibnamefont
  {Ott}},\ }\href {\doibase 10.1103/PhysRevC.96.065802} {\bibfield  {journal}
  {\bibinfo  {journal} {Phys. Rev. C}\ }\textbf {\bibinfo {volume} {96}},\
  \bibinfo {pages} {065802} (\bibinfo {year} {2017})}\BibitemShut {NoStop}%
\bibitem [{\citenamefont {Agrawal}\ \emph {et~al.}(2005)\citenamefont
  {Agrawal}, \citenamefont {Shlomo},\ and\ \citenamefont {Au}}]{KDE0v1}%
  \BibitemOpen
  \bibfield  {author} {\bibinfo {author} {\bibfnamefont {B.~K.}\ \bibnamefont
  {Agrawal}}, \bibinfo {author} {\bibfnamefont {S.}~\bibnamefont {Shlomo}}, \
  and\ \bibinfo {author} {\bibfnamefont {V.~K.}\ \bibnamefont {Au}},\ }\href
  {\doibase 10.1103/PhysRevC.72.014310} {\bibfield  {journal} {\bibinfo
  {journal} {Phys. Rev. C}\ }\textbf {\bibinfo {volume} {72}},\ \bibinfo
  {pages} {014310} (\bibinfo {year} {2005})}\BibitemShut {NoStop}%
\bibitem [{\citenamefont {Cao}\ \emph {et~al.}(2006)\citenamefont {Cao},
  \citenamefont {Lombardo}, \citenamefont {Shen},\ and\ \citenamefont
  {Giai}}]{LNS}%
  \BibitemOpen
  \bibfield  {author} {\bibinfo {author} {\bibfnamefont {L.~G.}\ \bibnamefont
  {Cao}}, \bibinfo {author} {\bibfnamefont {U.}~\bibnamefont {Lombardo}},
  \bibinfo {author} {\bibfnamefont {C.~W.}\ \bibnamefont {Shen}}, \ and\
  \bibinfo {author} {\bibfnamefont {N.~V.}\ \bibnamefont {Giai}},\ }\href
  {\doibase 10.1103/PhysRevC.73.014313} {\bibfield  {journal} {\bibinfo
  {journal} {Phys. Rev. C}\ }\textbf {\bibinfo {volume} {73}},\ \bibinfo
  {pages} {014313} (\bibinfo {year} {2006})}\BibitemShut {NoStop}%
\bibitem [{\citenamefont {Pais}\ and\ \citenamefont
  {Provid\^encia}(2016)}]{Pais16}%
  \BibitemOpen
  \bibfield  {author} {\bibinfo {author} {\bibfnamefont {H.}~\bibnamefont
  {Pais}}\ and\ \bibinfo {author} {\bibfnamefont {C.}~\bibnamefont
  {Provid\^encia}},\ }\href@noop {} {\bibfield  {journal} {\bibinfo  {journal}
  {Phys. Rev. C}\ }\textbf {\bibinfo {volume} {94}},\ \bibinfo {pages} {015808}
  (\bibinfo {year} {2016})}\BibitemShut {NoStop}%
\bibitem [{\citenamefont {Horowitz}\ and\ \citenamefont
  {Piekarewicz}(2001)}]{Horowitz01}%
  \BibitemOpen
  \bibfield  {author} {\bibinfo {author} {\bibfnamefont {C.~J.}\ \bibnamefont
  {Horowitz}}\ and\ \bibinfo {author} {\bibfnamefont {J.}~\bibnamefont
  {Piekarewicz}},\ }\href@noop {} {\bibfield  {journal} {\bibinfo  {journal}
  {Phys. Rev. Lett.}\ }\textbf {\bibinfo {volume} {86}},\ \bibinfo {pages}
  {5647} (\bibinfo {year} {2001})}\BibitemShut {NoStop}%
\bibitem [{\citenamefont {Fattoyev}\ \emph {et~al.}(2010)\citenamefont
  {Fattoyev}, \citenamefont {Horowitz}, \citenamefont {Piekarewicz},\ and\
  \citenamefont {Shen}}]{IUF}%
  \BibitemOpen
  \bibfield  {author} {\bibinfo {author} {\bibfnamefont {F.~J.}\ \bibnamefont
  {Fattoyev}}, \bibinfo {author} {\bibfnamefont {C.~J.}\ \bibnamefont
  {Horowitz}}, \bibinfo {author} {\bibfnamefont {J.}~\bibnamefont
  {Piekarewicz}}, \ and\ \bibinfo {author} {\bibfnamefont {G.}~\bibnamefont
  {Shen}},\ }\href {\doibase 10.1103/PhysRevC.82.055803} {\bibfield  {journal}
  {\bibinfo  {journal} {Phys. Rev. C}\ }\textbf {\bibinfo {volume} {82}},\
  \bibinfo {pages} {055803} (\bibinfo {year} {2010})}\BibitemShut {NoStop}%
\bibitem [{\citenamefont {Lalazissis}\ \emph {et~al.}(1997)\citenamefont
  {Lalazissis}, \citenamefont {K\"onig},\ and\ \citenamefont
  {Ring}}]{Lalazissis_PRC_1997}%
  \BibitemOpen
  \bibfield  {author} {\bibinfo {author} {\bibfnamefont {G.~A.}\ \bibnamefont
  {Lalazissis}}, \bibinfo {author} {\bibfnamefont {J.}~\bibnamefont {K\"onig}},
  \ and\ \bibinfo {author} {\bibfnamefont {P.}~\bibnamefont {Ring}},\ }\href
  {\doibase 10.1103/PhysRevC.55.540} {\bibfield  {journal} {\bibinfo  {journal}
  {Phys. Rev. C}\ }\textbf {\bibinfo {volume} {55}},\ \bibinfo {pages} {540}
  (\bibinfo {year} {1997})}\BibitemShut {NoStop}%
\bibitem [{\citenamefont {Todd-Rutel}\ and\ \citenamefont
  {Piekarewicz}(2005)}]{Todd-Rutel_PRL_2005}%
  \BibitemOpen
  \bibfield  {author} {\bibinfo {author} {\bibfnamefont {B.~G.}\ \bibnamefont
  {Todd-Rutel}}\ and\ \bibinfo {author} {\bibfnamefont {J.}~\bibnamefont
  {Piekarewicz}},\ }\href {\doibase 10.1103/PhysRevLett.95.122501} {\bibfield
  {journal} {\bibinfo  {journal} {Phys. Rev. Lett.}\ }\textbf {\bibinfo
  {volume} {95}},\ \bibinfo {pages} {122501} (\bibinfo {year}
  {2005})}\BibitemShut {NoStop}%
\bibitem [{\citenamefont {Toki}\ \emph {et~al.}(1995)\citenamefont {Toki},
  \citenamefont {Hirata}, \citenamefont {Sugahara}, \citenamefont {Sumiyoshi},\
  and\ \citenamefont {Tanihata}}]{TM1A}%
  \BibitemOpen
  \bibfield  {author} {\bibinfo {author} {\bibfnamefont {H.}~\bibnamefont
  {Toki}}, \bibinfo {author} {\bibfnamefont {D.}~\bibnamefont {Hirata}},
  \bibinfo {author} {\bibfnamefont {Y.}~\bibnamefont {Sugahara}}, \bibinfo
  {author} {\bibfnamefont {K.}~\bibnamefont {Sumiyoshi}}, \ and\ \bibinfo
  {author} {\bibfnamefont {I.}~\bibnamefont {Tanihata}},\ }\href {\doibase
  https://doi.org/10.1016/0375-9474(95)00161-S} {\bibfield  {journal} {\bibinfo
   {journal} {\npa}\ }\textbf {\bibinfo {volume} {588}},\ \bibinfo {pages}
  {c357} (\bibinfo {year} {1995})}\BibitemShut {NoStop}%
\bibitem [{\citenamefont {{Burgio}}\ \emph
  {et~al.}(2011{\natexlab{a}})\citenamefont {{Burgio}}, \citenamefont
  {{Ferrari}}, \citenamefont {{Gualtieri}},\ and\ \citenamefont
  {{Schulze}}}]{Burgio11}%
  \BibitemOpen
  \bibfield  {author} {\bibinfo {author} {\bibfnamefont {G.~F.}\ \bibnamefont
  {{Burgio}}}, \bibinfo {author} {\bibfnamefont {V.}~\bibnamefont {{Ferrari}}},
  \bibinfo {author} {\bibfnamefont {L.}~\bibnamefont {{Gualtieri}}}, \ and\
  \bibinfo {author} {\bibfnamefont {H.-J.}\ \bibnamefont {{Schulze}}},\ }\href
  {\doibase 10.1103/PhysRevD.84.044017} {\bibfield  {journal} {\bibinfo
  {journal} {\prd}\ }\textbf {\bibinfo {volume} {84}},\ \bibinfo {eid} {044017}
  (\bibinfo {year} {2011}{\natexlab{a}})}\BibitemShut {NoStop}%
\bibitem [{\citenamefont {{Paschalidis}}\ \emph {et~al.}(2012)\citenamefont
  {{Paschalidis}}, \citenamefont {{Etienne}},\ and\ \citenamefont
  {{Shapiro}}}]{Paschalidis12}%
  \BibitemOpen
  \bibfield  {author} {\bibinfo {author} {\bibfnamefont {V.}~\bibnamefont
  {{Paschalidis}}}, \bibinfo {author} {\bibfnamefont {Z.~B.}\ \bibnamefont
  {{Etienne}}}, \ and\ \bibinfo {author} {\bibfnamefont {S.~L.}\ \bibnamefont
  {{Shapiro}}},\ }\href {\doibase 10.1103/PhysRevD.86.064032} {\bibfield
  {journal} {\bibinfo  {journal} {\prd}\ }\textbf {\bibinfo {volume} {86}},\
  \bibinfo {eid} {064032} (\bibinfo {year} {2012})}\BibitemShut {NoStop}%
\bibitem [{\citenamefont {{Marques}}\ \emph {et~al.}(2017)\citenamefont
  {{Marques}}, \citenamefont {{Oertel}}, \citenamefont {{Hempel}},\ and\
  \citenamefont {{Novak}}}]{Marques17}%
  \BibitemOpen
  \bibfield  {author} {\bibinfo {author} {\bibfnamefont {M.}~\bibnamefont
  {{Marques}}}, \bibinfo {author} {\bibfnamefont {M.}~\bibnamefont {{Oertel}}},
  \bibinfo {author} {\bibfnamefont {M.}~\bibnamefont {{Hempel}}}, \ and\
  \bibinfo {author} {\bibfnamefont {J.}~\bibnamefont {{Novak}}},\ }\href
  {\doibase 10.1103/PhysRevC.96.045806} {\bibfield  {journal} {\bibinfo
  {journal} {\prc}\ }\textbf {\bibinfo {volume} {96}},\ \bibinfo {eid} {045806}
  (\bibinfo {year} {2017})}\BibitemShut {NoStop}%
\bibitem [{\citenamefont {{Lalit}}\ \emph {et~al.}(2019)\citenamefont
  {{Lalit}}, \citenamefont {{Mamun}}, \citenamefont {{Constantinou}},\ and\
  \citenamefont {{Prakash}}}]{Lalit19}%
  \BibitemOpen
  \bibfield  {author} {\bibinfo {author} {\bibfnamefont {S.}~\bibnamefont
  {{Lalit}}}, \bibinfo {author} {\bibfnamefont {M.~A.~A.}\ \bibnamefont
  {{Mamun}}}, \bibinfo {author} {\bibfnamefont {C.}~\bibnamefont
  {{Constantinou}}}, \ and\ \bibinfo {author} {\bibfnamefont {M.}~\bibnamefont
  {{Prakash}}},\ }\href {\doibase 10.1140/epja/i2019-12670-1} {\bibfield
  {journal} {\bibinfo  {journal} {\epja}\ }\textbf {\bibinfo {volume} {55}},\
  \bibinfo {eid} {10} (\bibinfo {year} {2019})}\BibitemShut {NoStop}%
\bibitem [{\citenamefont {Ishizuka}\ \emph {et~al.}(2008)\citenamefont
  {Ishizuka}, \citenamefont {Ohnishi}, \citenamefont {Tsubakihara},
  \citenamefont {Sumiyoshi},\ and\ \citenamefont {Yamada}}]{Ishizuka08}%
  \BibitemOpen
  \bibfield  {author} {\bibinfo {author} {\bibfnamefont {C.}~\bibnamefont
  {Ishizuka}}, \bibinfo {author} {\bibfnamefont {A.}~\bibnamefont {Ohnishi}},
  \bibinfo {author} {\bibfnamefont {K.}~\bibnamefont {Tsubakihara}}, \bibinfo
  {author} {\bibfnamefont {K.}~\bibnamefont {Sumiyoshi}}, \ and\ \bibinfo
  {author} {\bibfnamefont {S.}~\bibnamefont {Yamada}},\ }\href {\doibase
  10.1088/0954-3899/35/8/085201} {\bibfield  {journal} {\bibinfo  {journal}
  {{J. Phys. G}}\ }\textbf {\bibinfo {volume} {35}},\ \bibinfo {pages} {085201}
  (\bibinfo {year} {2008})}\BibitemShut {NoStop}%
\bibitem [{\citenamefont {{Burgio}}\ \emph
  {et~al.}(2011{\natexlab{b}})\citenamefont {{Burgio}}, \citenamefont
  {{Schulze}},\ and\ \citenamefont {{Li}}}]{Burgio2011}%
  \BibitemOpen
  \bibfield  {author} {\bibinfo {author} {\bibfnamefont {G.~F.}\ \bibnamefont
  {{Burgio}}}, \bibinfo {author} {\bibfnamefont {H.-J.}\ \bibnamefont
  {{Schulze}}}, \ and\ \bibinfo {author} {\bibfnamefont {A.}~\bibnamefont
  {{Li}}},\ }\href {\doibase 10.1103/PhysRevC.83.025804} {\bibfield  {journal}
  {\bibinfo  {journal} {\prc}\ }\textbf {\bibinfo {volume} {83}},\ \bibinfo
  {eid} {025804} (\bibinfo {year} {2011}{\natexlab{b}})}\BibitemShut {NoStop}%
\bibitem [{\citenamefont {Oertel}\ \emph {et~al.}(2016)\citenamefont {Oertel},
  \citenamefont {Gulminelli}, \citenamefont {Provid\^encia},\ and\
  \citenamefont {Raduta}}]{Oertel_EPJA_2016}%
  \BibitemOpen
  \bibfield  {author} {\bibinfo {author} {\bibfnamefont {M.}~\bibnamefont
  {Oertel}}, \bibinfo {author} {\bibfnamefont {F.}~\bibnamefont {Gulminelli}},
  \bibinfo {author} {\bibfnamefont {C.}~\bibnamefont {Provid\^encia}}, \ and\
  \bibinfo {author} {\bibfnamefont {A.~R.}\ \bibnamefont {Raduta}},\ }\href
  {\doibase 10.1140/epja/i2016-16050-1} {\bibfield  {journal} {\bibinfo
  {journal} {Eur. Phys. J. A}\ }\textbf {\bibinfo {volume} {52}},\ \bibinfo
  {pages} {50} (\bibinfo {year} {2016})}\BibitemShut {NoStop}%
\end{thebibliography}%

\end{document}